\documentclass[NETN,manuscript]{stjour-arxiv}

\makeatletter
\renewcommand\NAT@open{(}
\renewcommand\NAT@close{)}
\makeatother

\usepackage{hyperref}

\usepackage{subcaption}
\usepackage{float}
\usepackage{tikz}
\usetikzlibrary{shapes.geometric}
\usetikzlibrary{arrows.meta,positioning}
\usepackage{subcaption}
\usepackage[utf8]{inputenc}
\usepackage{csquotes}
\usepackage{textcomp}
\usepackage{xcolor}

\usepackage{ragged2e} 

\articletype{Research}

\nolinenumbers

\begin{document}

\justifying

\title{The Genetic and Environmental Architecture of the Human Functional Connectome}



\author[]
{Tanu Raghav\affil{1,2}, Daniel Guerrero\affil{1,2}, Uttara Tipnis\affil{3}, Julie Sara Benny\affil{1,2,4}, Mintao Liu\affil{1,2}, Mario Dzemidzic\affil{5}, Arian Ashourvan\affil{6}, Alex P. Miller\affil{7}, Beau Ances\affil{8}, Jaroslaw Harezlak\affil{9}, Joaqu\'{i}n Go\~{n}i\affil{1,2,10,*}}


\affiliation{1}{Edwardson School of Industrial Engineering, Purdue University, West Lafayette, IN, USA}
\affiliation{2}{Purdue Institute for Integrative Neuroscience, Purdue University, West Lafayette, IN, USA}
\affiliation{3}{Lawrence Livermore National Laboratory, Livermore, CA, USA}
\affiliation{4}{Department of Computational Science and Humanities, Indian Institute of Information Technology Kottayam, India}   

\affiliation{5}{Department of Neurology, Indiana University School of Medicine, Indiana Alcohol Research Center, Indianapolis, IN, USA}
\affiliation{6}{Department of Psychology, University of Kansas,  Lawrence, KS, USA}
\affiliation{7}{Department of Psychiatry, Indiana University School of Medicine, Indianapolis, IN, USA}
\affiliation{8}{Department of Neurology, Washington University in Saint Louis, School of Medicine, St Louis, MO, USA}
\affiliation{9}{Department of Epidemiology and Biostatistics, Indiana University, Bloomington, IN, USA}
\affiliation{10}{Weldon School of Biomedical Engineering, Purdue University, West Lafayette, IN, USA}

\vspace{0.1in}
\noindent {\footnotesize $^*$ Correspondence: Joaqu\'{i}n Go\~{n}i (jgonicor@purdue.edu).}

\correspondingauthor{Joaqu\'{i}n Go\~{n}i}{jgonicor@purdue.edu}


\keywords{Functional connectivity, fMRI conditions, Classical twin model, Multilayer community detection, heritability, environment}


\begin{abstract}
Functional connectivity varies across individuals due to genetic and environmental factors, yet classical twin models typically confound non-shared environment with measurement error and are largely limited to resting-state analyses.
We hypothesized that: i) explicitly modeling measurement error from repeated fMRI sessions enables more accurate application of classical twin models (ACE/ADE) to functional connectivity; ii) model applicability depends on scan-length and parcellation granularity; iii) genetic and environmental effects on functional connectomes show differentiated functional modules across conditions.
We extended ACE/ADE models to include a repeated-scan derived error term by analyzing monozygotic and dizygotic twins from the Young-Adult Human Connectome Project dataset. Genetic and environment variance components were estimated for all functional couplings across resting-state and task conditions, integrated across conditions using a minimum-error criterion, and analyzed using multilayer community detection across resolution scales.
Functional couplings segregated into distinct categories characterized by shared environmental, additive, dominant, or epistatic influences, with a substantial fraction not meeting twin-model assumptions. Integrating across conditions revealed hierarchical community structure in genetic and environmental components observed across community resolution scales. Incorporating measurement error into twin models improves interpretability and applicability at the functional connectome level, revealing that genetic and environmental influences are structured into coherent, multiscale brain networks.

\end{abstract}

\begin{authorsummary}
\justifying
Functional magnetic resonance imaging (fMRI) studies have shown that individuals exhibit distinctive patterns of brain connectivity that are reproducible across sessions, often referred to as “connectome fingerprints” \citep{Finn2015,Abbas2023}. Twin studies further indicate that these patterns are shaped by both genetic and environmental factors \citep{Colclough2017, Teeuw2019, Tassi2023, Pourmotabbed2024}. However, a key challenge has been disentangling true individual differences from session-to-session variability and measurement noise, which can obscure interpretation of genetic and environmental contributions.

In this study, we analyze twin fMRI data while explicitly accounting for measurement error derived from repeated scans. This approach increases the number of functional couplings that can be reliably analyzed and yields more accurate estimates of genetic and environmental effects. We also show that methodological choices (e.g., scan duration and brain parcellation granularity) substantially influence the reliability and interpretability of these estimates. By integrating results across resting-state and multiple task conditions, we demonstrate that genetic and environmental influences are not randomly distributed but instead organize into cohesive, hierarchical brain networks.

Overall, our framework provides a principled and practical approach to improve reliability in classical twin models applied to brain connectomics and offers new insights into how genes and environment shape large-scale brain function.
\end{authorsummary}

\section{Introduction}

Twin studies provide a principled framework to quantify the genetic and environmental contributions to inter-individual variability in human traits, including brain structure and function \citep{Boomsma2002,Polderman2015,twinchapter}. By contrasting monozygotic (MZ) and dizygotic (DZ) twins, classical variance component models partition phenotypic variance into additive genetic (A), dominant genetic (D), shared environmental (C), and non-shared environmental (E) components. These models have been widely used to estimate heritability across behavioral, cognitive, and neuroimaging phenotypes \citep{Craig2020,Hagenbeek2023,VanDongen2012}.

A key assumption of classical twin models is that measurement error is negligible and can therefore be subsumed within the non-shared environment component (E). While this assumption may be reasonable for highly controlled phenotypes (e.g., anthropometric measures), it is not tenable for functional connectivity (FC) derived from fMRI. This concern has been previously highlighted in other domains, including nutritional assessment \citep{Ulijaszek1999} and digital anthropometry using 3D optical measurements \citep{Mocini2023}, where measurement error can substantially bias inference. FC estimates are inherently affected by multiple sources of variability, including physiological noise, head motion, scanner instabilities, preprocessing choices, scan length and parcellation granularity. Consequently, the E component conflates true individual-specific environmental influences with measurement-related variability, limiting interpretability and introducing bias into variance decomposition \citep{Neale1992,Ge2017,Zuo2014}.

Functional connectivity (FC) derived from fMRI BOLD time series is inherently subject to multiple sources of variability that extend beyond underlying neural differences, distinguishing it from more stable phenotypic traits \citep{Murphy2013,Smith2011}. Measurement error arises from physiological fluctuations, head motion, and scanner- and sequence-related instabilities (e.g., drift, thermal noise), as well as sampling limitations such as short scan duration, low signal-to-noise ratio, and temporal autocorrelation \citep{Greve2013,Liu2016,Marta2009}. Additional variability is introduced by analytical choices, including preprocessing pipelines and their associated degrees of freedom, nuisance regression strategies (e.g., with or without global signal regression), parcellation granularity, temporal filtering, and the choice of FC estimator, all of which can systematically alter connectivity estimates even when applied to the same data. Beyond these technical factors, FC also exhibits state-dependent variability related to fluctuations in arousal, alertness, compliance, and ongoing cognition, leading to differences across repeated measurements within the same individual \citep{Wang2016,Gu2019,Chang2016,Laumann2017}.

As a result, FC variability reflects a composite of measurement noise and genuine within-subject fluctuations, yielding reliability that is heterogeneous across both individual functional connections and large-scale network organization. This complexity has motivated a parallel line of work on connectome fingerprinting, which quantifies the extent to which whole-brain FC patterns retain subject-specific structure across sessions and conditions, thereby isolating their trait-like component amid multiple sources of variability \citep{Finn2015,Horien2019,Abbas2023}. Building on this perspective, we explicitly model this variability by estimating edgewise (functional couplings) measurement error across repeated acquisitions, enabling a principled quantification of reliability at the level of the whole functional connectome.

\begin{figure}[H]
	\centering
        \fbox{ \includegraphics[width=1\textwidth,trim={1.5 0cm 1.5 0cm}, clip]{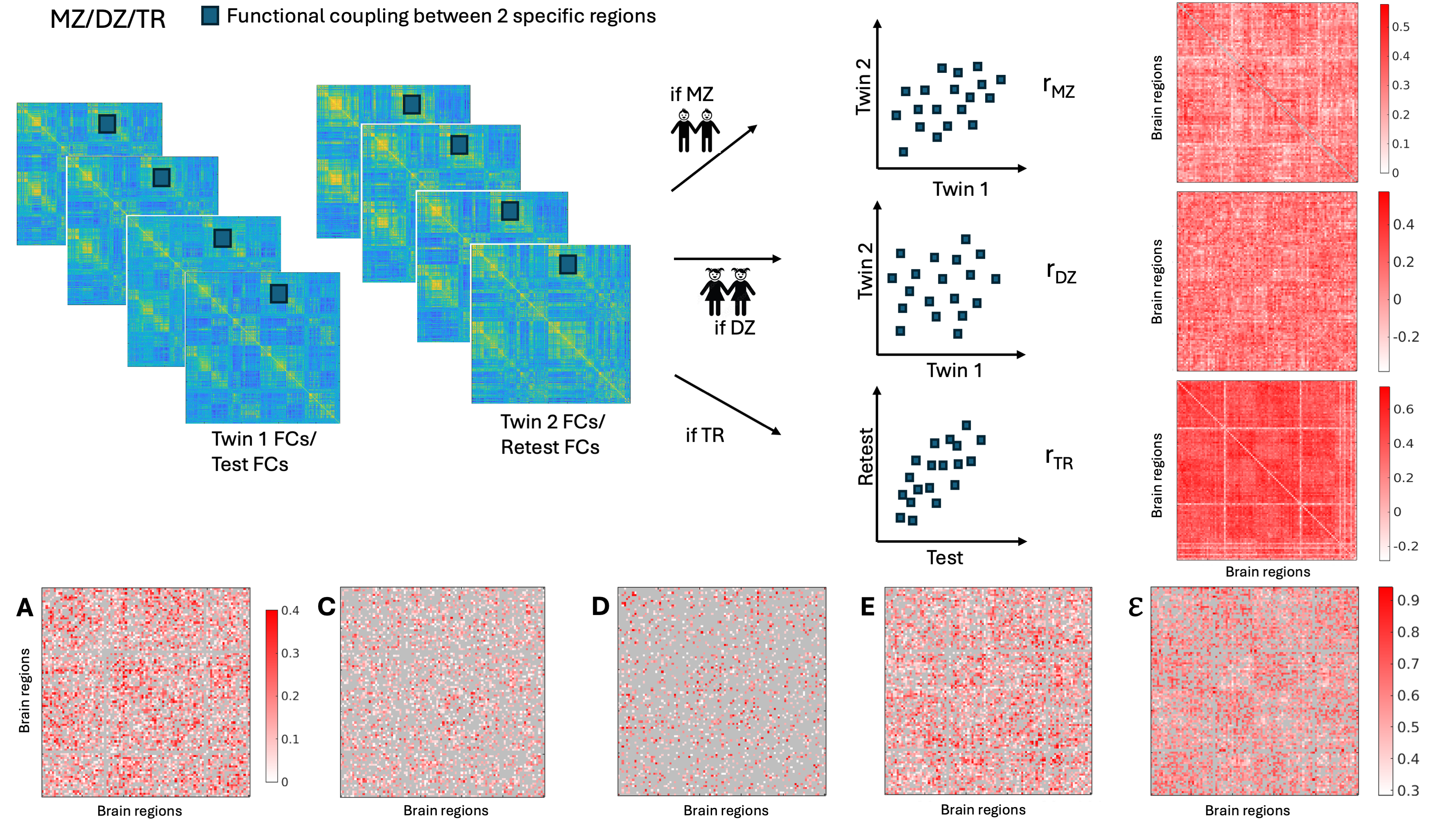}}
	\caption{\textbf{Workflow:} Functional coupling is extracted for twin $1$ and twin $2$ for all the MZ and DZ twin pairs. Correlation ($r_{MZ}$, $r_{DZ}$, $r_{TR}$) is computed on the extracted couplings from each FC, thus computing $A$, $C$, $D$, $E$ and $\mathcal{E}$ (see equations at Methods Section 2.5). The process is repeated for each coupling. This method is performed for all fMRI conditions resulting in eight (rest + $7$ tasks) different matrices of $A$, $C$, $D$, $E$ and $\mathcal{E}$. $r_{TR}$ refers to the correlation between functional couplings acquired from Test scan and Retest scan.}
    
	\label{schematic}
\end{figure}

Individual differences in FC have also been examined using twin designs, which probe the reliability and heritability of functional networks in both resting-state and task fMRI \citep{Abbas2023,Yao2023,Jansen2015, Pol2004,Pourmotabbed2024,Tassi2023}. Twin-model approaches applied to FC and related network features consistently indicate that portions of the connectome are heritable, often emphasizing additive genetic effects (A), while comparatively fewer studies explicitly characterize contributions from shared environment (C), dominant genetic effects (D), or unique environment (E) \citep{Glahn2010,Colclough2017,Teeuw2019, Blokland2008,Wu2024,Dominguez2018}.
However, most prior work does not explicitly model measurement error, instead subsuming it within the non-shared environment term (E), thereby inflating estimates of unique environmental contributions. \\
Test–retest designs (here, repeated scans) offer a means to disentangle session-level variability—arising from participant state and scanner-related factors—from stable subject-level connectivity patterns \citep{Zuo2014}. Accordingly, recent studies have begun to leverage repeated measurements to separate within-subject fluctuations from more stable components when estimating the heritability of resting-state connectivity, underscoring the importance of repeated acquisitions for more accurate interpretation of A, C, D, and E \citep{Ge2017, Chen2025}. In this context, explicitly quantifying measurement error provides a critical step toward disentangling true biological variability from acquisition- and session-driven noise in connectome-based heritability analyses.
Here, we extend the classical twin framework to explicitly incorporate measurement error estimated from test–retest fMRI data (see Figure~\ref{schematic} for illustrative diagram). Using monozygotic and dizygotic twins with repeated scans from the Young Adult Human Connectome Project dataset, we estimate variance components at the level of individual functional couplings across resting-state and task conditions. We further evaluate how scan design, including parcellation granularity and scan length, influences the proportion of couplings for which ACE/ADE models are applicable. To integrate results across conditions, we construct whole-brain variance component matrices using a minimum-error criterion and examine their organization using multilayer community detection across resolution scales. Together, this framework enables a functional connectome-wide characterization of genetic, environmental, and measurement-related contributions and their organization into large-scale brain networks.
We hypothesize that (i) explicit modeling of measurement error improves the reliability and interpretability of variance component estimates; (ii) the applicability of ACE/ADE models depends on scan design and parcellation granularity; and (iii) genetic and environmental influences on functional connectivity exhibit hierarchical modular organization across conditions. 

\section{Methods}
\subsection{Human Connectome Project Young-Adult dataset}

\begin{table}[htbp]
\centering
\begin{tabular}{lcccc}
\hline
\textbf{Group} & \textbf{Sex (M/F)} & \textbf{Age (mean $\pm$ SD)} & \textbf{Education (mean $\pm$ SD)} \\
\hline
Unrelated & 223/203 & 28.67 $\pm$ 3.78 & 14.99 $\pm$ 1.77 \\
MZ        & 132/100 & 29.27 $\pm$ 3.34 & 14.98 $\pm$ 1.83 \\
DZ        & 70/56   & 28.69 $\pm$ 3.43 & 15.29 $\pm$ 1.62 \\
\hline
\end{tabular}
\caption{Demographic characteristics of the study sample across unrelated individuals, monozygotic (MZ), and dizygotic (DZ) twins. Values are presented as counts or mean $\pm$ standard deviation. There were no significant differences across groups in sex distribution ($p=0.51$), age ($p=0.11$), or years of education ($p=0.22$).}
\label{tab:demographics}
\end{table}

In this work, we used data from the Human Connectome Project Young Adult dataset ($1,200$ participants) \citep{HCP2013}. All procedures were approved by the Institutional Review Board (IRB) of Washington University in St. Louis, and all participants provided written informed consent prior to participation. We extracted three different subsets. The first consists of $426$ unrelated participants ($223$ women, mean age: $28.67$ years old, range: $22\mathord{-}36$), while the remaining subsets include $116$ pairs of Monozygotic (MZ) twins and $63$ pairs of Dizygotic (DZ) twins.  All DZ pairs of twins were of the same sex, to minimize differences in shared environment between the MZ and DZ cohorts \citep{twinchapter}. 
fMRI data included both resting state (RS) and seven tasks: emotion processing (EM), gambling (GAM), language (LAN), motor (MOT), relational processing (REL), social cognition (SOC), and working memory (WM). For simplicity, we refer to the resting state and the tasks as fMRI conditions, or simply conditions.
For each condition, participants underwent two sessions corresponding to two different acquisitions (left-to-right or LR, and right-to-left or RL). The resting-state scans (``REST$1$'' and ``REST$2$'') were acquired on two different days for a total of four sessions. Only the two sessions from REST$1$ were utilized in this work.

\subsection{Brain parcellation granularity}
In this work, we have used a hierarchical collection of functional brain atlases of the cortex, known as the Schaefer parcellations \citep{Schaefer2018}, based on resting-state fMRI data from $1,489$ participants. 
The Schaefer parcellations used surface alignment but are available both in volumetric and grayordinate space and at ten granularity levels: $100\mathord{-}1000$ in steps of $100$. 
The Schaefer parcellations are available both in volumetric and grayordinate space. Since the grayordinate versions of the parcellations are in the same surface space as the HCP fMRI data, it is straightforward to map them onto the fMRI data. Furthermore, the alignment between the fMRI data and the Schaefer parcellations is improved when using surface-mapping. Thus, we used the surface-based mapping to map the $100\mathord{-}900$ granularity parcellations onto the fMRI data. At the time of the data processing for this study, we could not map the $1000$ Schaefer parcellation successfully for the HCP Young Adult dataset. 
\subsection{Preprocessing of the Young Adult HCP dataset}
A ``minimal'' preprocessing pipeline from the HCP was employed \citep{Glasser2013}, comprising artifact removal, motion correction, and registration to standard template. Full details can be found in earlier publications \citep{Glasser2013, Smith2013}.
We added the following steps to the ``minimal'' pipeline. For resting-state fMRI data we: (i) regressed out the global gray matter signal from the voxel time courses \citep{Power2014}, (ii) applied a first-order Butterworth bandpass filter in the forward and the reverse directions [$0.001\mathord{-}0.08Hz$ \citep{Power2014}; MATLAB functions butter and filtfilt], and (iii) z-scored and averaged, per brain regions, the voxel time courses, excluding any outlier time points falling outside three standard deviation from the mean (workbench software, wb\_command -cifti-parcellate). For task-fMRI, we performed the same steps, but applied a wider frequency range for the bandpass filter ($0.001\mathord{-}0.25Hz$) \citep{Amico2019}, since the optimal frequency range for task fMRI data is still unclear \citep{Cole2014}.

\subsection{Estimation of whole-brain functional connectomes}
Functional connectivity between two brain regions ($q_1$ and $q_2$) was estimated using Pearson’s correlation coefficient (MATLAB command corr):
\begin{equation}
    r=\sum_{i=1}^{m} \frac{(q_1(i)-\bar{q_1})(q_2(i)-\bar{q_2})}{\sigma_1\sigma_2}
\end{equation}
where\\
$r$ is the Pearson’s correlation coefficient.\\
$q_1(i)$ and $q_2(i)$ are the $i_{th}$ elements of $q_1$ and $q_2$, respectively.\\
$\bar{q_1}$ and $\bar{q_2}$ are the sample means of $q_1$ and $q_2$, respectively.\\
$\sigma_1$ and $\sigma_2$ are the standard deviations of $q_1$ and $q_2$, respectively.

Computing the correlations between all pairs of brain regions results into a symmetric correlation matrix of size $m\times m$, where $m$ is the number of brain regions for a given parcellation. We refer to this
object as a whole-brain functional connectome (FC). For each participant, we computed an FC for each of the two sessions (also referred to as test-retest), each fMRI condition (resting state and all tasks), and each parcellation granularity ($100\mathord{-}900$). 

\subsection{The classical twin model}
The classical twin model is a statistical framework that, for any phenotype, compares the similarity of MZ twins and DZ twins to partition the variance of a trait into additive and dominant genetics and shared and unique environment  \citep{twinchapter}. MZ twins are genetically identical while DZ twins share about $50\%$ of their genetic material. In both cases, it is expected that MZ and DZ twins have a large portion of shared environment. By comparing similarities within MZ and DZ twins, the classical twin model partitions the phenotype variance of a trait into the following four components: i) Additive genetic effects ($A$) represent variance attributable to the cumulative, linear effects of individual alleles inherited from parents; ii) Shared environmental effects ($C$) capture variance arising from environmental factors that make twins raised together more similar to each other; iii) Dominant genetic effects ($D$) represent variance due to non-additive interactions between alleles at the same genetic locus. iv) Non-shared environmental effects ($E$) capture variance due to environmental influences unique to each individual, notably including measurement error. Because this is a partition of phenotypic variance, the sum of all four components equals to one.


Correlation between monozygotic (MZ) twins can be attributed to additive genetic effects ($A$), shared environment ($C$), and dominant genetic effects ($D$), such that the expected correlation can be expressed as $r_{MZ} = A + C + D$. In contrast, dizygotic (DZ) twins share on average $50\%$ of their additive genetic variance and $25\%$ of their dominant genetic variance, along with shared environmental factors, yielding the expression $r_{DZ} = 0.5A + C + 0.25D$ \citep{twinchapter}.   

 \begin{figure}[h]
 \centering





\includegraphics[width=0.8\textwidth]{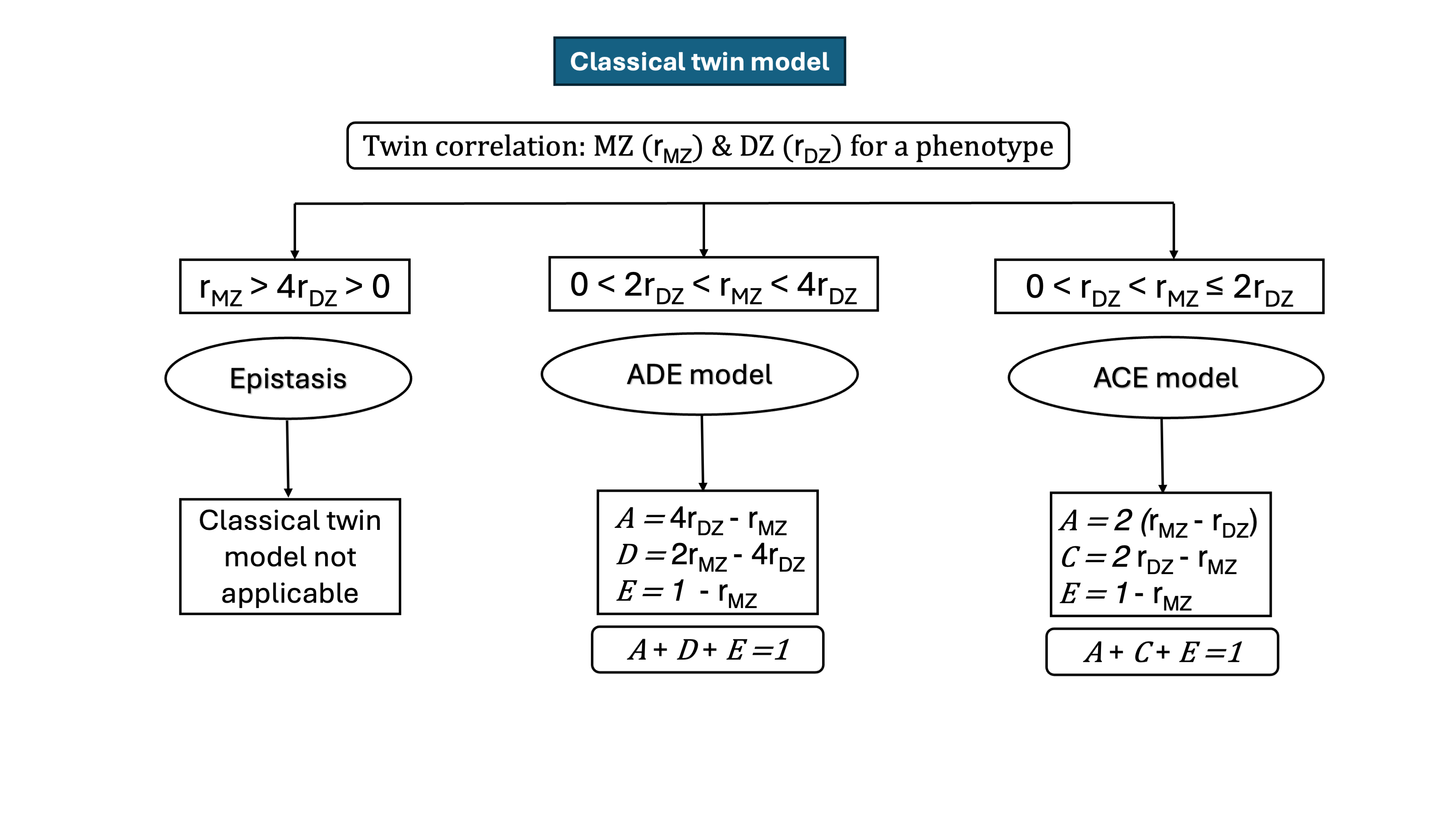}
\vspace{-1cm}
 \caption{Model selection (ACE vs ADE) in the Classical Twin Model based on comparison between MZ and DZ twin correlations ($r_{MZ}$ and $r_{DZ}$ respectively. When the condition $r_{DZ} \geq 0.5r_{MZ}$ is not met, systematic use of ACE model leads to inflated additive heritability estimations.}
 \label{ctm_selection}
 \end{figure}

This leads to a set of three equations and 4 unknowns:
\[
r_{MZ} = A + C + D
\]
\[
r_{DZ} = 0.5A + C + 0.25D 
\]
\[
A + C + D + E = 1
\]

\begin{figure}
	\centering
        \includegraphics[width=1\textwidth]{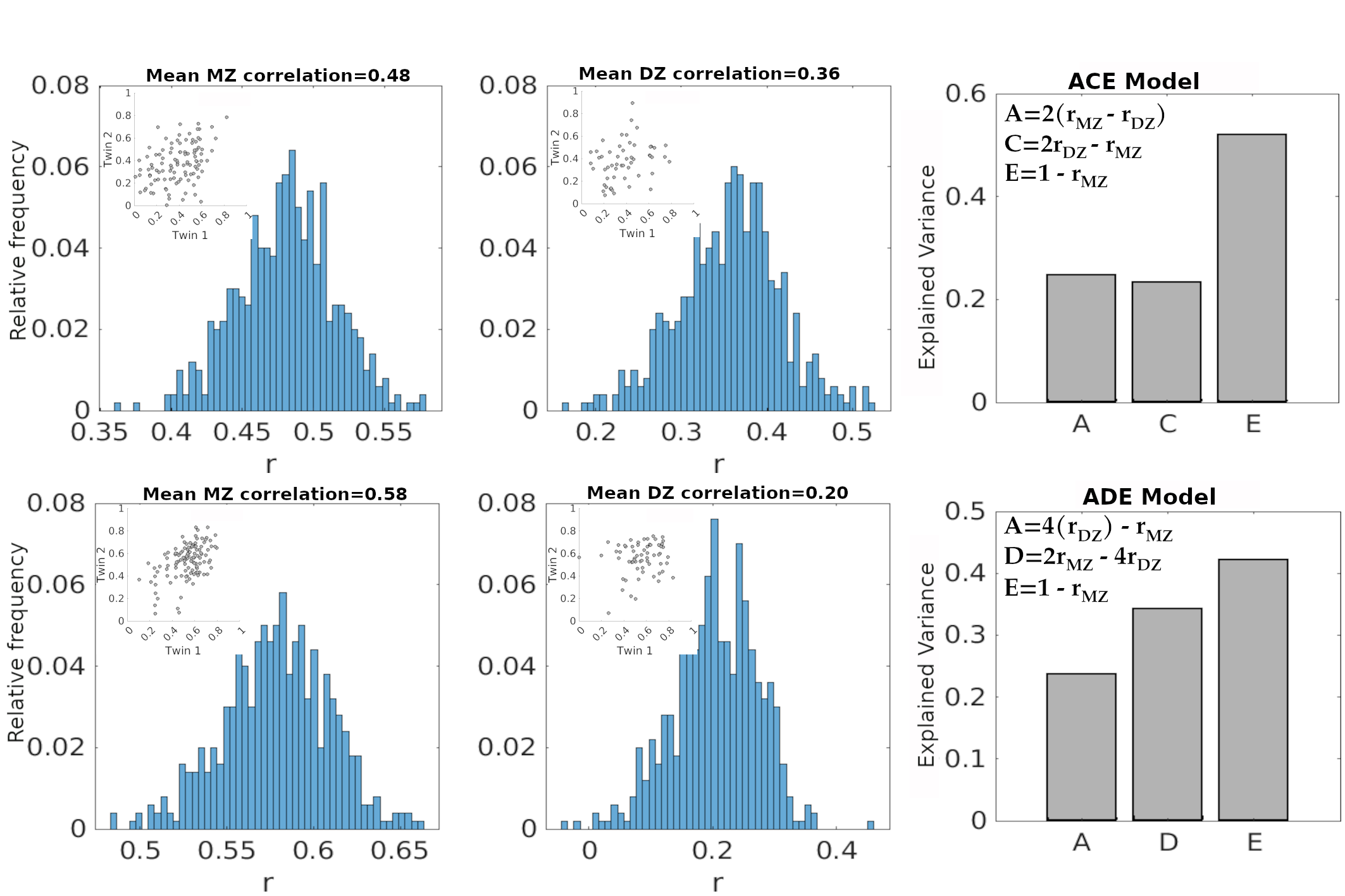}
	\caption{Estimation of variance components using twin correlations: Histograms show sampling distributions of Pearson correlation coefficients (r) for monozygotic (MZ) and dizygotic (DZ) twin pairs, obtained via repeated resampling for two representative functional couplings during resting state. Insets show scatter plots for a single illustrative sampling instance, depicting twin 1 versus twin 2 values for MZ and DZ pairs. 
    In the top row, the sampled correlations ($r_{MZ} = 0.48, r_{DZ} = 0.36$) satisfy the ACE model criteria, yielding variance component estimates of additive genetics ($A = 2(r_{MZ}-r_{DZ}) = 0.24$), shared environment ($C = 2r_{DZ}-r_{MZ} = 0.24$), and unique environment ($E = 0.52$). In the bottom row, the sampled correlations ($r_{MZ} = 0.58, r_{DZ} = 0.2$) satisfy the ADE model criteria, yielding $A = 0.24$, dominant genetic effects ($D = 0.34$), and $E = 0.42$. Bar plots summarize the corresponding variance component estimates for each example.}
	\label{scatterplot}
\end{figure}

Because the classical twin model includes four variance components but only three independent equations, it is not possible to estimate all components simultaneously. To address this limitation, the model estimates either $C$ or $D$, as these terms are negatively confounded \citep{twinchapter, Tassi2023}. Shared environmental effects ($C$) contribute similarly to monozygotic (MZ) and dizygotic (DZ) twins, tending to increase both $r_{MZ}$ and $r_{DZ}$, while dominant genetic effects ($D$) disproportionately increase $r_{MZ}$ relative to $r_{DZ}$. Consequently, when $0 < r_{\mathrm{DZ}} < r_{\mathrm{MZ}} \leq 2r_{\mathrm{DZ}}$, the ACE model is selected, implying negligible $D$. In contrast, when $0 < 2r_{\mathrm{DZ}} < r_{\mathrm{MZ}} < 4r_{\mathrm{DZ}}$, the ADE model is selected, implying negligible $C$ \citep{twinchapter}. A schematic summary of this selection procedure is shown in Figure~\ref{ctm_selection}. Figure~\ref{scatterplot} further illustrates, for a representative functional coupling, the estimation of $r_{MZ}$ and $r_{DZ}$ and the corresponding variance component decomposition.

 In consequence, the ACE model equations are as follows:

\[
\begin{aligned}
A &= 2(r_{MZ}-r_{DZ}) \\
C & =2r_{DZ}-r_{MZ} \\
E &= 1 - r_{MZ}
\end{aligned}
\]
such that 
\[
    A+C+E=1
\]
Analogously, the ADE equations are as follows:
\[
\begin{aligned}
A &= 4r_{DZ} - r_{MZ} \\
D &= 2r_{MZ} - 4r_{DZ} \\
E &= 1 - r_{MZ}
\end{aligned}
\]
such that 
\[
    A+D+E=1
\]

Our first step consisted of estimating the above-mentioned factors for traits. In our study, those are the functional couplings. Therefore, for each functional coupling, and for MZ and DZ twin pairs separately, we computed the Pearson's correlation coefficient between twin $1$ and twin $2$ in each pair. As a result, we obtain zygosity specific correlations $r_{MZ}$ and $r_{DZ}$. 
In order to obtain reliable estimations of $r_{MZ}$ and $r_{DZ}$, we implemented an iterative sampling procedure. For each iteration: i) $80\%$ of the twin-pairs (MZ or DZ) were randomly selected. ii) for each pair of twins, labels 'twin1' and 'twin2' were randomly assigned. iii) Pearson's correlation coefficients between 'twin1' and 'twin2' functional couplings were obtained. This process was repeated $500$ times (Figure~\ref{convergence_corr}), producing a distribution of correlation coefficients. The final correlation estimate was obtained as the mean of this distribution and was then used to estimate $A$, $C$, $D$ and $E$ for each coupling using above equations (Figure~\ref{schematic})

\subsection{Community detection on ACE/ADE components using Multiplex Networks.}
\label{sec:comm_det}
In this study, we use the concept of multiplex networks to assess the organization into communities of each variance component of ACE/ADE models across different resolutions. A multiplex network consists of $m$ layers sharing a common set of $N$ nodes, where each layer encodes a distinct type of interaction \citep{Kivela2014,Domenico2016}. Formally, the network is represented as
$\mathcal{M} = (G, E)$, where $G = \{ g_a \mid a = 1, \ldots, m \}$ denotes the set of layers, and each layer $g_a = (N, E_a)$ contains the same node set and a layer-specific set of intra-layer edges. Inter-layer connectivity is defined by edges $e_{ab}$ linking corresponding mirror nodes across layers, such that intra-layer edges ($a=b$) capture layer-specific interactions, while inter-layer edges ($a \neq b$) encode couplings between interaction types. We consider ordinal multiplex networks, in which inter-layer connections exist only between adjacent layers and only between mirror nodes \citep{Mucha2010}. The resulting supra-adjacency matrix has a block structure with intra-layer adjacency matrices $A^{a}$ on the diagonal and inter-layer couplings $\omega I$ on the off-diagonal blocks. 

For example, adjacency matrix $A$ for $m$-layer multiplex network can be written as;
\[
\mathbf{A} =
\begin{pmatrix}
A^{1}      & \omega I & 0          & \cdots     & 0 \\
\omega I   & A^{2}    & \omega I   & \cdots     & 0 \\
0          & \omega I & A^{3}      & \ddots     & \vdots \\
\vdots     & \vdots   & \ddots     & \ddots     & \omega I \\
0          & 0        & \cdots     & \omega I  & A^{m}
\end{pmatrix},
\]
where $A^{a}$ is the $N \times N$ adjacency matrix corresponding to layer $a$, $I$ denotes the identity matrix, and $\omega$ represents the inter-layer coupling strength.

Communities describe groups of nodes more strongly connected to each other than to nodes outside of their community, as compared to a null model representing what would be expected in a network of similar size and density with randomly distributed connections \citep{Newman2006}. Modularity describes a formal measurement of this \citep{NewmanandGirvan2004}, which can be used in iterative processes of finding a partition of a network with an optimal set of community assignments. In the current study the GenLouvain community detection algorithm is used \citep{Mucha2010}, which involves the following multiplex modularity measurement that is formulated to work on multilayer networks:
\begin{equation*}
    Q=\frac{1}{2\mu}\sum_{ijsr}[(A_{ijs}-\gamma_s\frac{k_{is}k_{js}}{2m_s})\delta_{sr}+\delta_{ij}\omega_{jsr}]\delta(g_{is}, g_{jr})
\end{equation*}
Here $\mu$ is the total edge weight of the network, $A_{ijs}$ is the adjacency matrix between nodes $i$ and $j$ at layer $s$. $\frac{k_{is}k_{js}}{2m_s}$ describes the matrix of expected weights under the null model. The resolution parameter $\gamma_s$ sets the weight of intralayer edges, and $\omega_{jsr}$ sets the weight of inter layer edges. $g$ refers to community assignments: $g_{is}$ and $g_{jr}$ are the community assignments of node $i$ in layer $s$, and node $j$ in layer $r$, respectively. $\delta$ describes the Kronecker delta, which for $\delta(g_{is}, g_{jr})$ is $1$ if $is = jr$, and $0$ if $is \neq jr$.
 GenLouvain community detection algorithm was applied to $A$, $C$, $E$ and $\mathcal{E}$ multiplex networks. 

\begin{figure}[H]
    \centering   
    \includegraphics[width=1\textwidth]{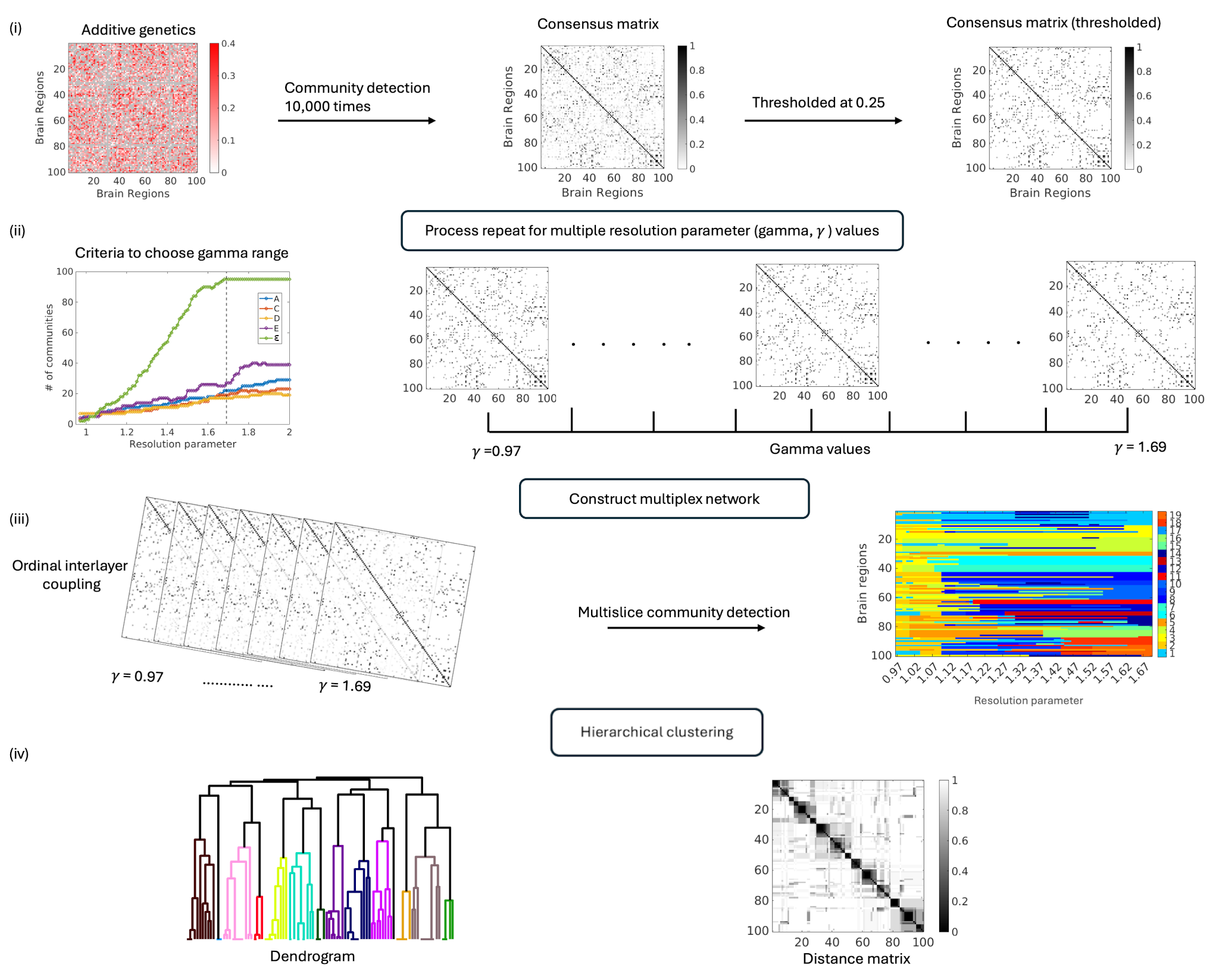}
\caption{Multilayer consensus community detection pipeline: Community detection is repeated across a range of resolution parameters ($\gamma$) to generate consensus matrices summarizing stable node co-assignments. Thresholded consensus matrices are treated as layers of a multiplex network with ordinal interlayer coupling, and multislice community detection is applied to identify communities coherent across resolutions. Hierarchical clustering of multislice partitions reveals nested modular organization. }
\label{Flowchart_Hierarchy}
\end{figure}

To identify stable community structure across multiple resolutions, we implemented a multilayer consensus community detection approach (Figure~\ref{Flowchart_Hierarchy}). For a given matrix (A, C, D, E, $\mathcal{E}$), we vary the resolution parameter $\gamma$ across a predefined range. As shown in Figure~\ref{num_comm}, the number of detected communities increases with $\gamma$, with component-specific stabilization patterns. In particular, the measurement error ($\mathcal{E}$) component exhibits a clear plateau at $\gamma = 1.69$, which we use to define the upper bound of the resolution range for the multilayer network. The lower bound is chosen as the smallest $\gamma$ at which all components exhibit at least two communities. For each $\gamma$ value, the Louvain algorithm is executed 10,000 times to account for its stochasticity. The resulting partitions are then combined into a consensus matrix, in which each entry represents the proportion of runs in which a given pair of nodes is assigned to the same community \citep{Fortunato2012}. To ensure robustness, the consensus matrix is thresholded at $0.25$, retaining only node pairs that co-occurred in at least 25$\%$ of runs. The thresholded matrix, termed as the agreement matrix, captures the stable co-assignment structure at that specific $\gamma$ value. Repeating this procedure across all $\gamma$ values produces a series of agreement matrices, each representing the network’s modular organization at a different resolution.
These agreement matrices are then treated as layers in a multiplex network, where each layer corresponding to a particular $\gamma$ value, and inter-layer coupling preserves the identity of nodes across layers. Multilayer community detection is then performed on this multilayer structure to extract communities that are coherent across resolutions \citep{Mucha2010}. Finally, a hierarchical clustering procedure is applied to the multislice partitions to reveal nested modular organization, yielding a hierarchy of communities that captures the network’s structure from fine to coarse grained levels (see Fig.~\ref{Flowchart_Hierarchy}).

\subsection{Comparing partitions via Adjusted Rand Index} \label{sec:ARI}

To quantify the similarity between community partitions derived from variance component connectivity matrices and functional network organization \citep{Yeo2011}, we employed the Adjusted Rand Index (ARI). ARI is a measure of similarity between two partitions that evaluates the consistency of pairwise assignments of elements to clusters while accounting for agreement expected under random labeling \citep{Hubert1985}. Mathematically, 
\begin{equation}
\mathrm{ARI}=
\frac{\mathrm{Index} - \mathbb{E}[\mathrm{Index}]}
     {\mathrm{Max\ Index} - \mathbb{E}[\mathrm{Index}]} 
\end{equation}
where Index denotes the observed Rand index between two partitions, 
$\mathbb{E}[\mathrm{Index}]$ is the expected agreement under random labeling, 
and $\mathrm{Max\ Index}$ is the maximum possible agreement, equal to 1.

In the present study, ARI was used to compare community labels obtained from the additive genetic (A), common environment (C), dominant genetic (D), unique environment (E), and measurement error ($\mathcal{E}$) components across the seven Yeo functional network labels. This approach enables an assessment of whether variance component specific modular organization aligns with known functional network structure, independent of the number or size of modules. ARI values close to zero indicate weak correspondence, whereas negative values indicate agreement lower than expected by chance.

\begin{table}[]
\centering
\begin{tabular}{p{6cm}|c|c|c|c|c|c|c}
\hline
Trait & $r_{\mathrm{MZ}}$ & $r_{\mathrm{DZ}}$ & A & C & D & E & Model \\
\hline
Height                          & 0.94 & 0.76 & 0.35 & 0.58 & -- & 0.06 & ACE \\
Weight                          & 0.77 & 0.52 & 0.56 & 0.23 & -- & 0.19 & ACE \\
BMI                             & 0.65 & 0.32 & 0.63 & -- & 0.01 & 0.35 & ADE \\
Episodic Memory                 & 0.50 & 0.07 & -- & -- & -- & -- & Epistasis \\
Executive Function -- Cognitive & 0.43 & 0.13 & 0.22 & -- & 0.02 & 0.75 & ADE \\
Executive Function -- Inhibition& 0.44 & 0.09 & -- & --   & -- & -- & Epistasis \\
Fluid Intelligence              & 0.52 & 0.34 & 0.66 & 0.01 & -- & 0.32 & ACE \\
Language Reading/Decoding       & 0.70 & 0.49 & 0.37 & 0.31 & -- & 0.32 & ACE \\
Processing Speed                & 0.33 & 0.07 & -- & --   & -- & -- & Epistasis \\
ASR Internalizing Raw Score     & 0.41 & 0.11 & 0.03 & -- & 0.37 & 0.58 & ADE \\
ASR Externalizing Raw Score     & 0.40 & 0.06 & --   & -- & --  & --    & Epistasis \\
\hline
\end{tabular}
\caption{For each trait, the correlations between monozygotic ($r_{MZ}$) and dizygotic ($r_{DZ}$) twin pairs are reported, along with the estimated variance components attributable to additive genetic effects (A), common environmental effects (C), dominant genetic effects (D), and unique environmental effects (E). The only applicable model (ACE, ADE, or epistasis-dominant patterns) is indicated in the final column based on standard twin-model assumptions and observed correlation patterns. Dashes indicate components not included or not identifiable under the selected model. 
}
\label{tabACDE}
\end{table}

\section{Results}
\subsection{Heritable vs. Environmental Proportion in Phenotypic Traits}

To contextualize the heritability estimates of functional connectivity, we quantified genetic and environmental contributions to a set of demographic, cognitive, and behavioral traits in twins \citep{HCP2013}. Cognitive measures in the Human Connectome Project (HCP) are primarily derived from the NIH Toolbox Cognition Battery, a standardized and validated framework for assessing multiple domains of cognitive function across the lifespan \citep{Gershon2013, Weintraub2013}. Specifically, we examined height, weight, fluid intelligence, episodic memory, processing speed, language reading/decoding, and executive function (including cognitive flexibility and inhibition). Fluid intelligence was assessed using the Penn Matrix Reasoning Test (PMAT24), which measures abstract reasoning and problem-solving ability independent of acquired knowledge. Episodic memory was measured using the Picture Sequence Memory Test, which evaluates the ability to encode and recall sequences of visual information. Processing speed was assessed using the Pattern Comparison Processing Speed Test, reflecting the efficiency of visual information processing. Executive function was evaluated using tasks such as the Dimensional Change Card Sort and Flanker Task, which measure cognitive flexibility and inhibitory control, respectively, while language ability was assessed using the Oral Reading Recognition Test, capturing reading and decoding skills. In addition, behavioral and emotional functioning was assessed using the Achenbach System of Empirically Based Assessment (ASR), where internalizing scores reflect anxiety/depressive and withdrawn symptom dimensions, and externalizing scores reflect rule-breaking and aggressive behavior profiles \citep{Achenbach2003}.

For each trait, we computed Pearson correlation coefficients separately for monozygotic (MZ) and dizygotic (DZ) twin pairs, yielding zygosity-specific correlations ($r_{MZ}$ and $r_{DZ}$); Table \ref{tabACDE}). The distribution of all demographic, cognitive, and behavioral traits across the full sample of MZ and DZ twins is shown in Figure~\ref{trait_histogram}
. Classical twin modeling was applied only to traits satisfying $r_{MZ}>r_{DZ}>0$. When $2r_{DZ}<r_{MZ}<4r_{DZ}$, we assumed contributions from additive and dominant genetic factors along with unique environmental influences, and thus applied the ADE model, decomposing variance into additive genetic (A), dominant genetic (D), and unique environmental (E) components. Conversely, when $r_{DZ}<r_{MZ}<2r_{DZ}$, we applied the ACE model, partitioning variance into additive genetic (A), shared environmental (C), and unique environmental (E) components. Cases where $r_{MZ}>4r_{DZ}$ were not adequately captured by either model and were interpreted as indicative of epistatic (non-additive) genetic interactions  \citep{VanDongen2012, Tassi2023}.\\
Twin correlations revealed distinct genetic architectures across traits (Table \ref{tabACDE}). Height and weight exhibited strong heritability, with notable contributions from shared environmental factors. Among cognitive traits, fluid intelligence and language/reading decoding showed substantial additive genetic influences (A = $0.66$ and $0.37$, respectively), consistent with an ACE model. In contrast, executive function (inhibition) and processing speed were better explained by an ADE model, reflecting contributions from both additive and dominant genetic effects. Episodic memory displayed a marked discrepancy between $r_{MZ}=0.47$ and $r_{DZ}=0.07$, a pattern inconsistent with ACE or ADE assumptions and suggestive of epistatic interactions. ASR internalizing ($r_{MZ}=0.41$, $r_{DZ}=0.11$) and externalizing ($r_{MZ}=0.40$, $r_{DZ}=0.06$) scores similarly exhibited low DZ correlations relative to MZ; however, internalizing was consistent with an ADE model reflecting non-additive genetic effects, whereas externalizing was indicative of epistatic interactions.
Overall, demographic traits were predominantly influenced by additive genetic effects, whereas cognitive traits exhibited a more complex architecture involving additive, dominant, and epistatic genetic contributions. Full estimates of A, C, and E components for each trait are reported in Table \ref{tabACDE}.

\begin{figure}[H]
    \centering   
    \includegraphics[width=0.8\textwidth]{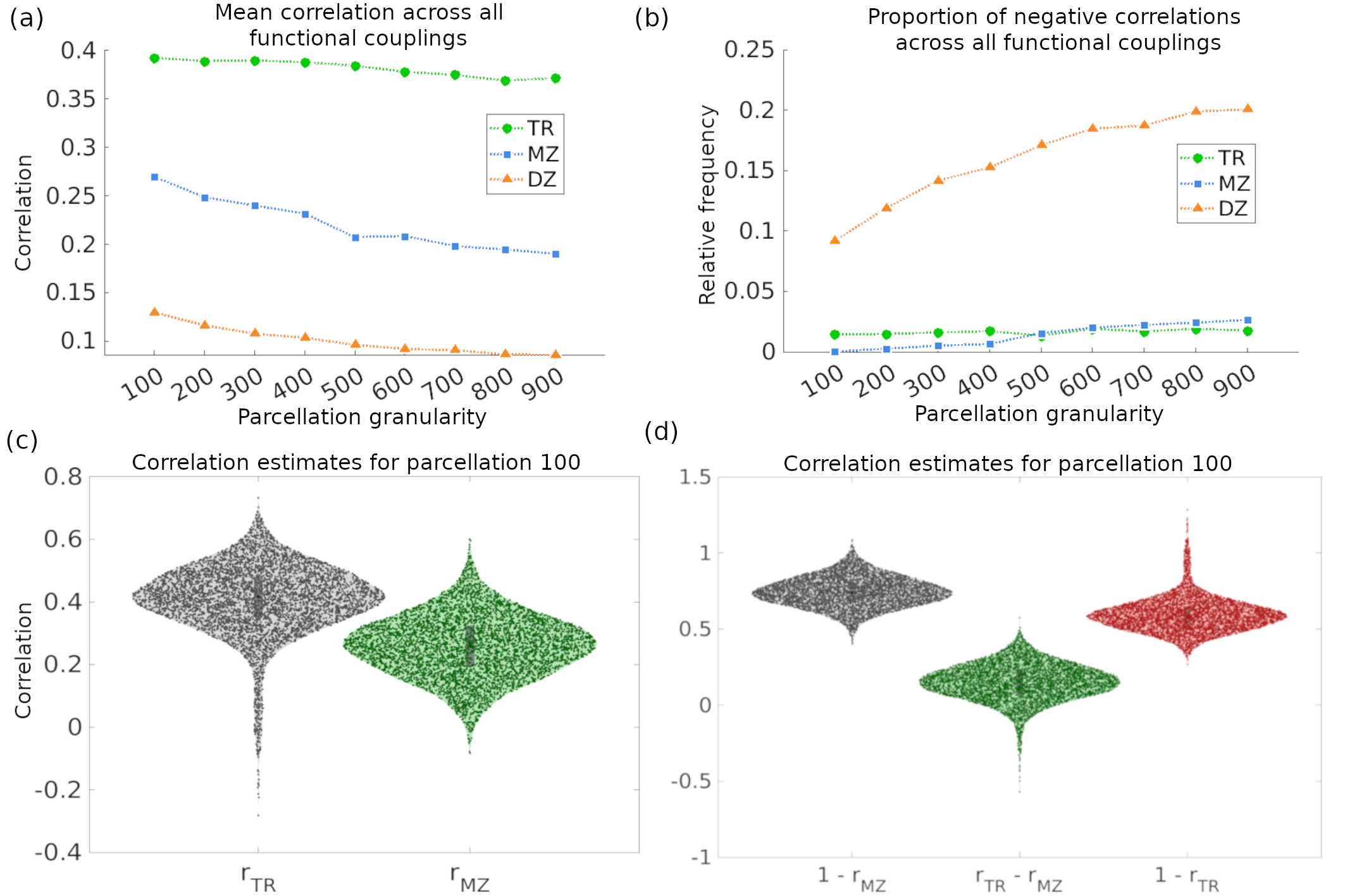}
\caption{(a) Mean edgewise correlations ($r_{TR}$, $r_{MZ}$, $r_{DZ}$) for resting-state as a function of parcellation granularity. 
(b) Proportion of negative correlations across all functional couplings for resting-state, shown for $r_{TR}$, $r_{MZ}$, and $r_{DZ}$ as a function of parcellation granularity. 
(c) Violin plots of test--retest reliability ($r_{TR}$) and monozygotic twin similarity ($r_{MZ}$) for resting-state at parcellation granularity 100. 
(d) Comparison of classical and extended estimates of non-shared environment for resting-state at parcellation granularity 100: classical ACE estimate ($1 - r_{MZ}$), corrected estimate from ACE$\mathcal{E}$ ($r_{TR} - r_{MZ}$), and measurement error ($1 - r_{TR}$).}
\label{violin}
\end{figure}

\subsection{Extension of ACE and ADE models}
In classical ACE and ADE frameworks, the non-shared environment component ($E$) is typically estimated as $1 - r_{MZ}$, where $r_{MZ}$ denotes the similarity between monozygotic (MZ) twins. This formulation conflates true individual-specific environmental variance with measurement-related variability. To disentangle these contributions, we extend the classical twin models by explicitly incorporating a measurement error term derived from within-subject test--retest data, yielding the ACE$\mathcal{E}$ and ADE$\mathcal{E}$ formulations. In this framework, test--retest reliability ($r_{TR}$) defines the upper bound of variance attributable to biological and environmental factors, while the residual variance, $1 - r_{TR}$, reflects measurement error arising from scanner noise, preprocessing variability, and other non-systematic sources. Accordingly, we redefine the non-shared environment as $E = r_{TR} - r_{MZ}$, capturing only reliable variance not explained by genetic or shared environmental effects, and explicitly model measurement error as $\mathcal{E} = 1 - r_{TR}$.

When evaluating the effect of parcellation granularity, we considered two complementary metrics: the mean correlations ($r_{TR}$, $r_{MZ}$, and $r_{DZ}$) across all functional couplings, and the proportion of negative correlations. The former reflects signal preservation, whereas the latter serves as an indicator of noise, as negative values are not expected for $r_{TR}$, $r_{MZ}$, or $r_{DZ}$. While test--retest reliability ($r_{TR}$) remains largely stable across parcellation granularities, the average monozygotic ($r_{MZ}$) and dizygotic ($r_{DZ}$) twin correlations decrease as granularity increases (Figure~5A). Concurrently, the proportion of negative correlations increases with finer parcellations, particularly for $r_{DZ}$, whereas $r_{TR}$ and $r_{MZ}$ remain comparatively stable (Figure~5B). Based on these observations in this sample, a parcellation of 100 regions was the best option to maximize mean correlations, while minimizing the presence of negative correlations.

Figure~\ref{violin}d illustrates the resulting variance decomposition under the ACE$\mathcal{E}$ model. Notably, the conventional estimate of non-shared environment ($1 - r_{MZ}$) systematically exceeds the corrected estimate ($r_{TR} - r_{MZ}$), indicating inflation due to unmodeled measurement error. By incorporating $r_{TR}$, the proposed framework separates true environmental effects from noise, providing a more interpretable partition of variance. Overall, this extension highlights that the traditional ACE/ADE models overestimates environmental contributions by attributing unreliable variance to $E$. In contrast, the ACE$\mathcal{E}$ formulation constrains interpretable variance through $r_{TR}$ and isolates measurement-related variability via $\mathcal{E}$. 

\begin{figure}[H]
    \centering
    \includegraphics[width=0.8\textwidth]{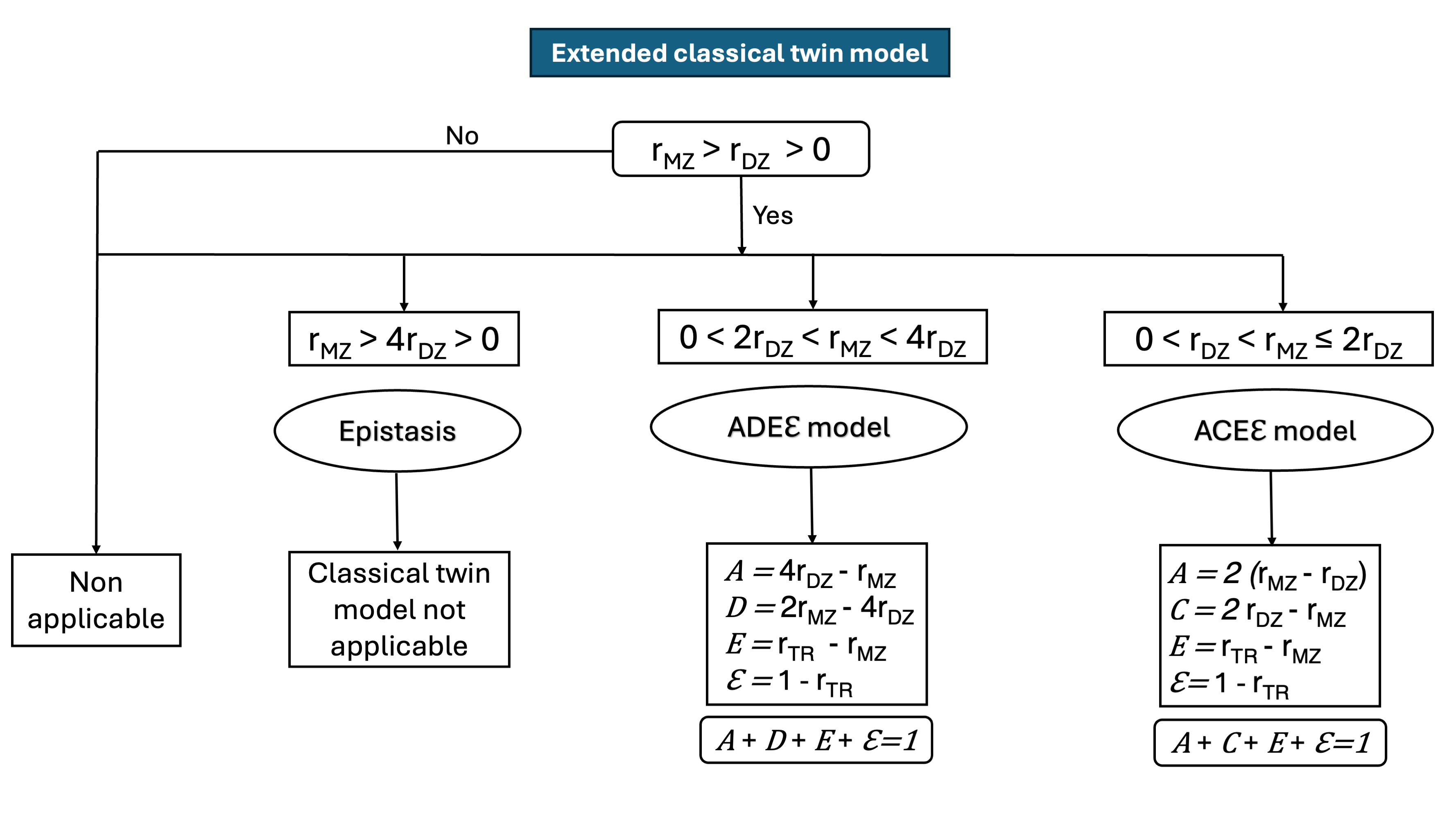}
    \caption{Flowchart illustrating the criteria for detecting applicable functional couplings under the ACE$\mathcal{E}$ and ADE$\mathcal{E}$ models based on monozygotic (MZ) and dizygotic (DZ) twin correlations. Depending on the relationships between $r_{MZ}$ and $r_{DZ}$, couplings are assigned to the ACE$\mathcal{E}$, ADE$\mathcal{E}$, or epistasis regimes. Each branch specifies the corresponding variance component estimates ($A$, $C$, $D$, $E$) and measurement error term ($\mathcal{E}$). This framework also identifies couplings for which classical twin modeling assumptions are not satisfied, and thus no variance decomposition is applied.}
    \label{flowchart}
\end{figure}

\subsection{Model-Based Categorization of Functional Couplings under Extended Twin Framework}

Building on the extended twin modeling framework, we next quantified the genetic and environmental contributions to individual functional couplings across all fMRI conditions. For each pairwise coupling, we computed the Pearson correlation between twin 1 and twin 2 separately for monozygotic (MZ) and dizygotic (DZ) pairs, yielding zygosity-specific correlations $r_{MZ}$ and $r_{DZ}$ (Figure~\ref{schematic}). This procedure was performed independently for each fMRI condition.

Only couplings satisfying the criterion $r_{MZ} > r_{DZ} > 0$ were retained for further analysis; all others were excluded and labeled as not applicable. Among the retained couplings, three distinct regimes emerged (Figure~\ref{flowchart}):

(1) For couplings satisfying $0 < r_{DZ} < r_{MZ} \leq 2r_{DZ}$, we applied the ACE$\mathcal{E}$ model, estimating additive genetic ($A$), shared environmental ($C$), unique environmental ($E$), and measurement error ($\mathcal{E}$) components.

(2) For couplings satisfying $0 < 2r_{DZ} < r_{MZ} < 4r_{DZ}$, we applied the ADE$\mathcal{E}$ model, estimating additive genetic ($A$), dominant genetic ($D$), unique environmental ($E$), and measurement error ($\mathcal{E}$) components.

(3) For couplings where $r_{MZ} > 4r_{DZ} > 0$, the assumptions underlying classical twin models are violated, precluding variance decomposition under ACE$\mathcal{E}$ or ADE$\mathcal{E}$. These cases are interpreted as reflecting non-additive gene–gene interactions (epistasis).

\begin{figure}
    \centering
     \includegraphics[width=\linewidth]{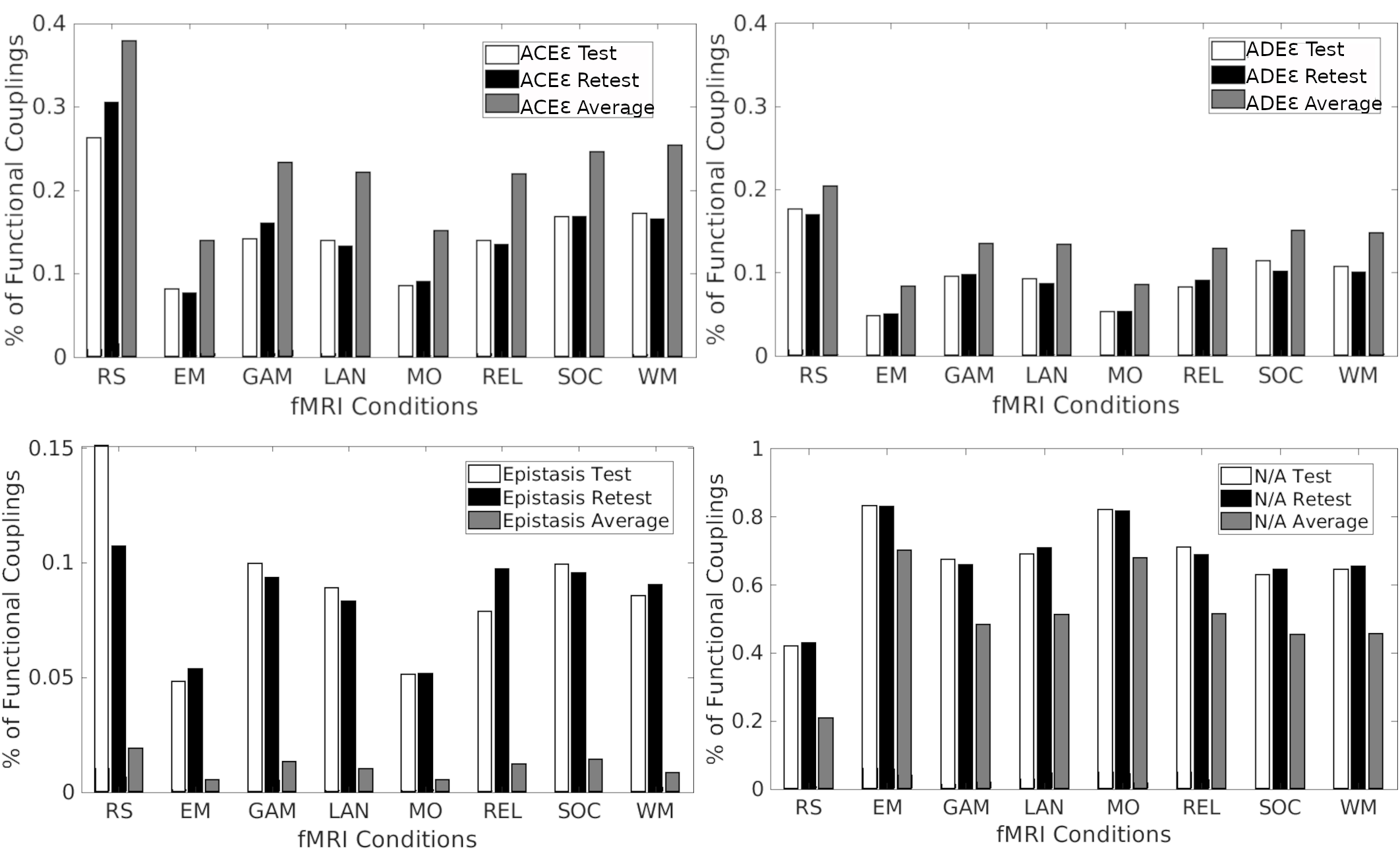}
    \caption{Functional couplings attributed to each variance component (A, C, D, E, error term $\mathcal{E}$, and epistasis) across fMRI conditions. ACE$\mathcal{E}$ and ADE$\mathcal{E}$ average results were restricted to functional couplings for which both test and retest data satisfied the same model classification (ACE$\mathcal{E}$ or ADE$\mathcal{E}$). }
    \label{variance_components}
\end{figure}

 The distribution of functional couplings attributed to $A$, $C$, $D$, $E$, and $\mathcal{E}$ across all fMRI conditions are summarized in Figure~\ref{variance_components}. Across both resting-state and task conditions, the ACE$\mathcal{E}$ model accounts for a larger fraction of couplings than the ADE$\mathcal{E}$ model, indicating that most functional couplings are primarily influenced by additive genetic and environmental factors rather than dominant genetic effects. Nevertheless, ACE$\mathcal{E}$ and ADE$\mathcal{E}$ explain comparable proportions of couplings across task conditions, suggesting that additive and dominant genetic influences contribute consistently across different brain states. Couplings classified as epistatic are also observed in both resting-state and task conditions, consistent with the presence of non-additive gene–gene interactions.

It is noteworthy that a substantial proportion of couplings are classified as not applicable (N/A), as they do not satisfy the heritability criterion ($r_{MZ} > r_{DZ} > 0$; see Figure~\ref{variance_components}). This fraction reaches up to 80\% in certain task conditions (e.g., Emotion and Motor), with most tasks exhibiting approximately 70\% non-applicable couplings, whereas the proportion is lowest for resting-state data (20\% approximately). This analysis was performed separately on test (white bars) and retest (black bars) sessions. The proportions of couplings across categories are highly consistent between sessions for all fMRI conditions, indicating robust estimates. However, the specific couplings classified as applicable or non-applicable show minor variability between sessions. To obtain more stable estimates, we averaged model outputs (A,C,D,E,$\mathcal{E}$) across test and retest sessions (gray bars in Figure~\ref{variance_components}). This aggregation was restricted to functional couplings for which both test and retest data satisfied the same model classification (ACE$\mathcal{E}$ or ADE$\mathcal{E}$). Under this constraint, averaging led to an increased number of couplings classified under the ACE$\mathcal{E}$ and ADE$\mathcal{E}$ models, while reducing the proportion of epistatic and non-applicable couplings across all conditions. All subsequent analyses are therefore based on these averaged estimates.

\begin{figure}
    \centering
   \begin{subfigure}[b]{0.5\textwidth}
        \centering
        \includegraphics[width=\textwidth]{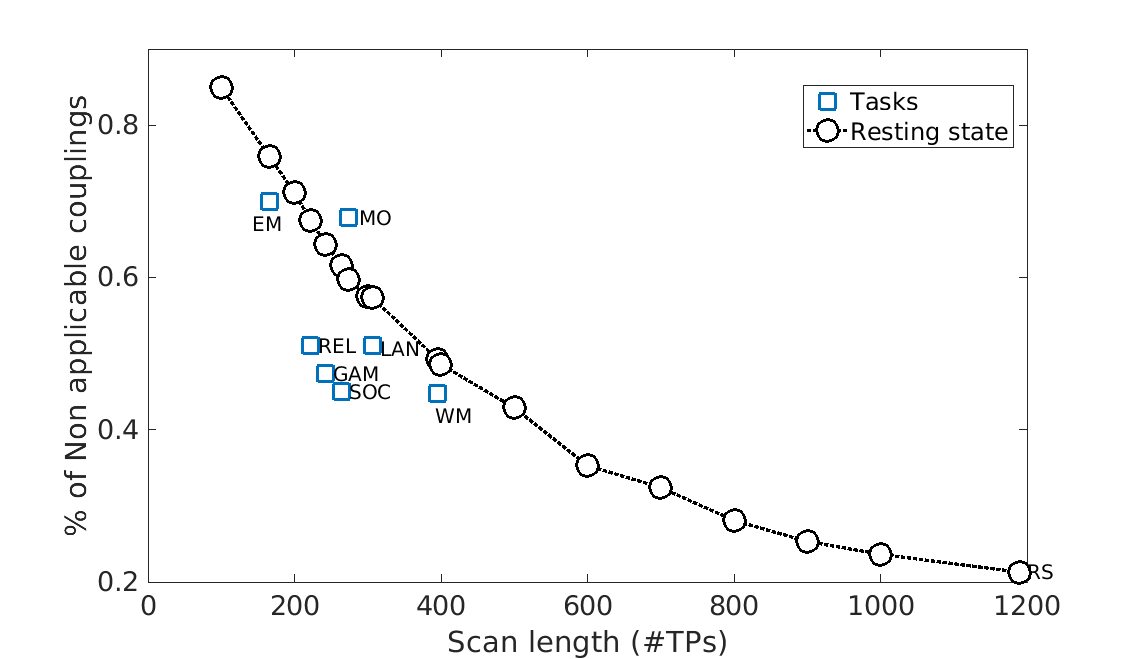}
    \end{subfigure}
    \begin{subfigure}[b]{0.465\textwidth}
        \centering
        \includegraphics[width=\textwidth]{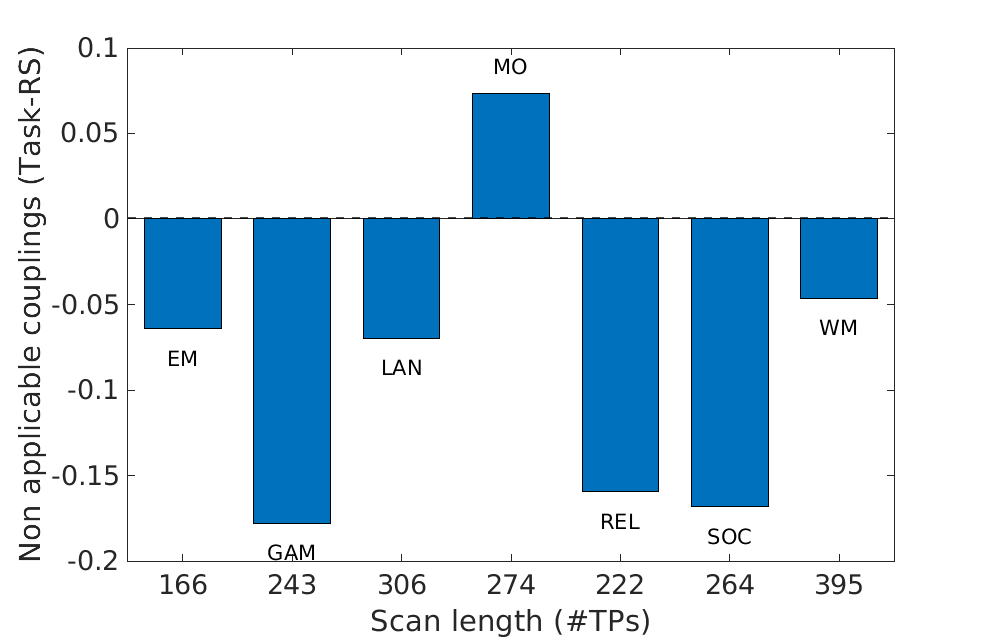}
    \end{subfigure}
    \caption{Left: Effect of scan length on the proportion of non-applicable couplings. The fraction of couplings excluded from the extended twin model decreases as scan length increases. Task conditions (blue squares) have shorter acquisitions (100–400 time points) and exhibit a higher proportion of non-applicable couplings, whereas resting-state (black circles), characterized by longer scan durations, shows substantially fewer excluded couplings. Right: Difference in the number of non-applicable couplings for task and resting state data at equivalent scan lengths.}
    \label{na_scan}
\end{figure}

\subsection{Distribution of Genetic and Environmental Variance Components along the Functional Connectome}

After computing $r_{MZ}$ and $r_{DZ}$ for each functional coupling and assigning couplings to the appropriate model, we estimated the corresponding variance components. For couplings satisfying the ACE$\mathcal{E}$ criteria, variance proportions were computed as $A = 2(r_{MZ} - r_{DZ})$, $C = 2r_{DZ} - r_{MZ}$, $E = r_{TR} - r_{MZ}$, and $\mathcal{E} = 1 - r_{TR}$. For couplings classified under the ADE$\mathcal{E}$ model, components were estimated as $A = 4r_{DZ} - r_{MZ}$, $D = 2r_{MZ} - 4r_{DZ}$, $E = r_{TR} - r_{MZ}$, and $\mathcal{E} = 1 - r_{TR}$. This procedure was applied to every functional coupling, yielding full-brain matrices (at each parcellation scale) for each variance component ($A$, $C$, $D$, $E$, and $\mathcal{E}$). As analyses were performed across both resting-state and task fMRI conditions, this resulted in eight matrices per component, one per condition.

Figure~\ref{schematic} illustrates the analysis workflow and presents the resulting component matrices for the resting-state condition, highlighting the distribution of genetic, environmental and measurement error related contributions across functional couplings. The matrix $A$ identifies couplings primarily driven by additive genetic effects, reflecting heritability. The matrix $C$ captures shared environmental influences common to both twins, while the matrix $D$ reflects dominant (non-additive) genetic effects. The matrix $E$ represents the variance attributable to unique environmental influences after accounting for reliability, and the matrix $\mathcal{E}$ quantifies the measurement error and residual noise not explained by twin similarity. Together, these component-specific networks provide a characterization of the genetic and environmental architecture underlying brain organization for the entire functional connectome.


\begin{figure}
    \centering
    
    \includegraphics[width=\linewidth]{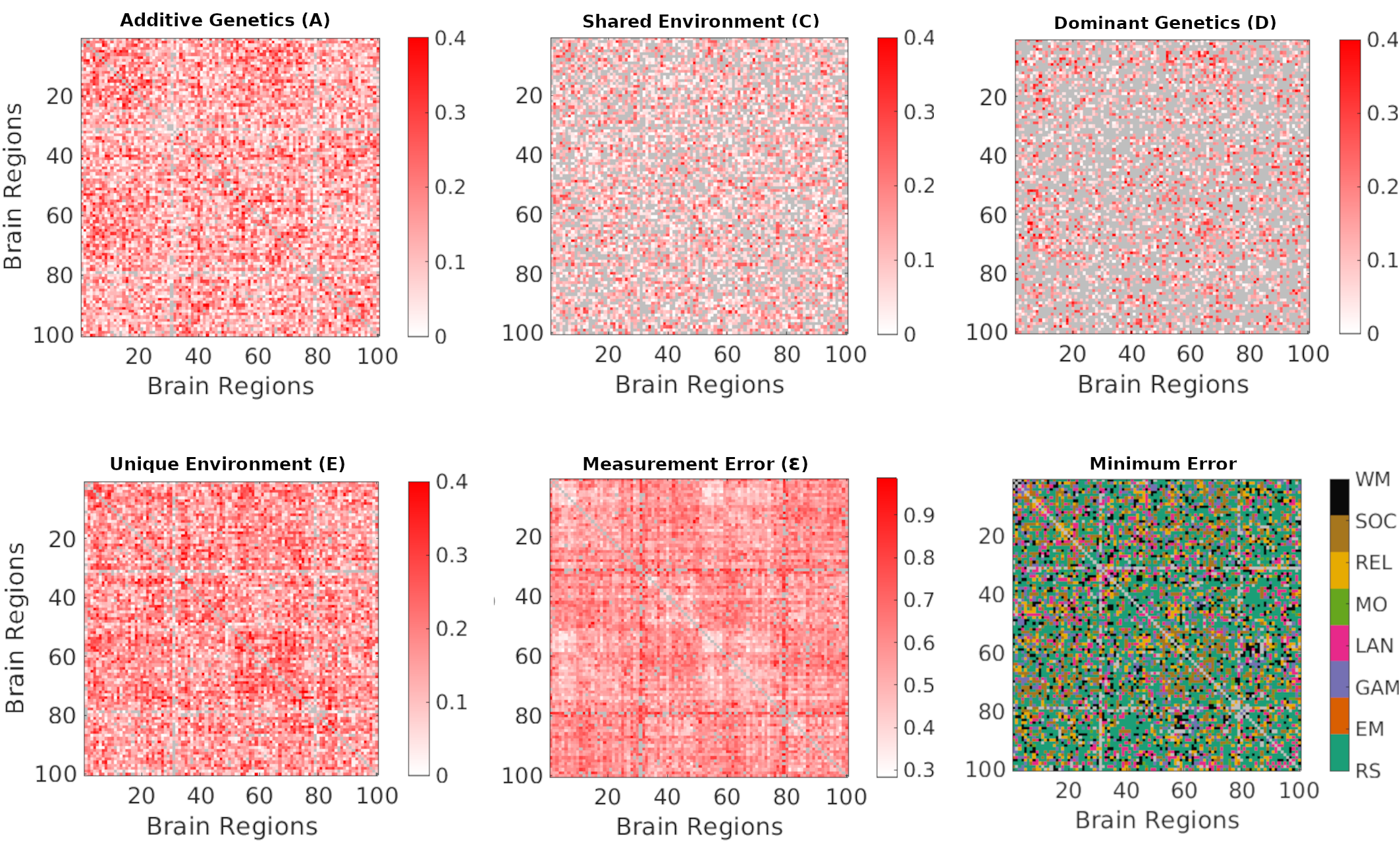}
    \caption{$A$, $C$, $D$, $E$, and measurement error ($\mathcal{E}$) matrices obtained by integrating all fMRI conditions at parcellation granularity 100. For each functional coupling, the estimate corresponding to the condition with the lowest $\mathcal{E}$ was selected as the most reliable value. Applying this minimum-error selection across all couplings yielded merged whole-brain matrices (A, C, D, E) that integrate information across conditions while minimizing measurement noise. The bottom-right panel shows the edge-wise distribution of fMRI conditions contributing the minimum-error estimates.
    }
    \label{fig:ACDE_mxd}
\end{figure}

\subsection{Integration Across fMRI Conditions based on minimum error}

Figure~\ref{variance_components} shows that the fraction of non-applicable (N/A) couplings varies across resting-state and task conditions. While such variability may partly reflect differences in brain state, scan length also differs substantially across fMRI conditions in the HCP Young Adult dataset. Specifically, resting-state scan has the most time points ($1190$), whereas task scans range from 166 to 395 time points (Emotion = 166, Gambling = 243, Language = 306, Motor = 274, Relational = 222, Social = 264, Working Memory = 395). Prior work has demonstrated that longer fMRI acquisitions improve the reliability and stability of functional connectivity estimates and enhance the interpretability of network-level analyses \citep{Birn2013, Ooi2025, Jarrahi2021, Wahab2022}.

To explicitly assess the effect of scan length, we quantified the proportion of N/A couplings in resting-state data while systematically truncating the time series in increments of 100 time points. In addition, we truncated resting-state data to match the duration of each task condition, enabling two complementary analyses: (i) characterization of the dependence of N/A proportion on scan length for resting-state; (ii) comparison of N/A proportions across resting-state and task conditions at matched scan lengths. As shown in Figure~\ref{na_scan}, the proportion of non-applicable couplings decreases with increasing scan length and differs between resting-state and task conditions. A multiple linear regression analysis confirmed that both scan length and condition (task vs. rest) are significant predictors reducing the percentage of non-applicable couplings ($p < 0.05$).
Consistent with this interpretation, truncating resting-state data to match task durations yields N/A proportions that closely resemble those observed in several task conditions—particularly Emotion, Motor, Language, and Working Memory. 

Given the relatively small fraction of functional couplings for which ACE/ADE models are applicable within individual fMRI conditions (ranging from $25\%$ to $38\%$ for ACE and $8\%$ to $21\%$ for ADE; Figure~\ref{variance_components}), we integrated results across conditions using a minimum-error approach. For each variance component (A, C, D, E, and $\mathcal{E}$), eight matrices were obtained corresponding to the eight fMRI conditions (resting state and seven tasks), yielding eight candidate estimates per functional coupling. We then selected, for each coupling, the estimate associated with the condition exhibiting the lowest measurement error ($\mathcal{E}$). Applying this procedure across all couplings produced combined A, C, D, and E matrices that retain, for each coupling, the most reliable estimate across conditions (Figure~\ref{fig:ACDE_mxd}). As shown in Figure~\ref{fig:ACDE_mxd}(f), resting-state data contribute the highest proportion of selected couplings (58\%), followed by Social (10\%) and Working Memory (9\%). This distribution reflects the greater reliability afforded by longer resting-state acquisitions, while still incorporating task-specific information when it provides more stable estimates.



\subsection{Organization into functional modules is driven by phenotypic decomposition}

\begin{figure}
	\centering
        \includegraphics[width=1\textwidth]{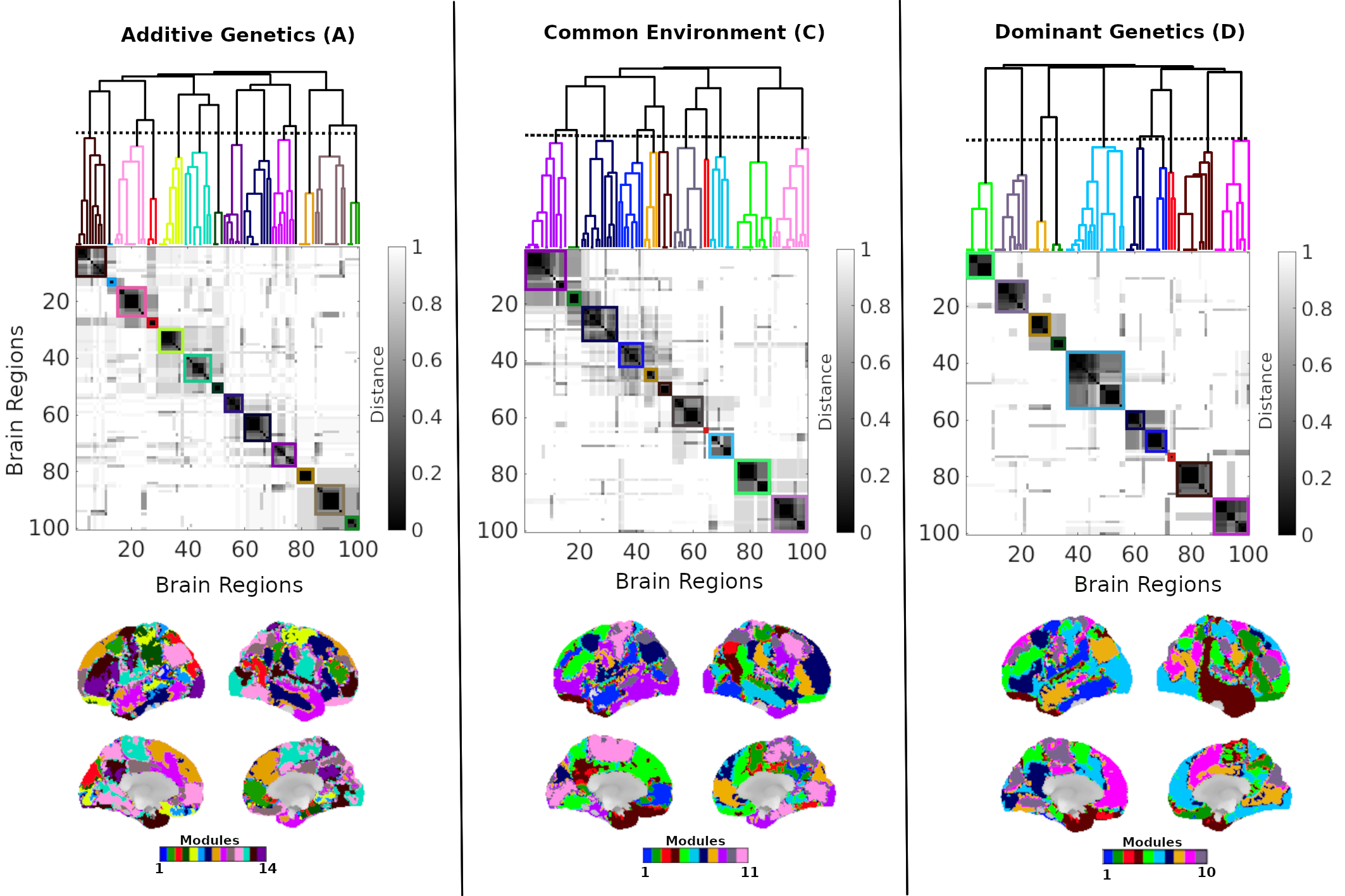}
	\caption{Hierarchical community structure of additive genetics (A), shared environmental (C), and dominant genetics (D) components of the whole-brain functional connectome. For each component, the top panels display dendrograms derived from hierarchical clustering of brain regions based on the pairwise similarity of their community assignment patterns across resolution parameters, together with the corresponding distance matrices. Colored blocks denote groups of regions forming stable communities. The bottom panels project these hierarchical modules onto the lateral and medial cortical surfaces, illustrating the spatial organization of each component. Colors indicate distinct modules within each component independently and are not comparable across components.
    }
	\label{A_Community_render}
\end{figure}

To characterize the hierarchical organization of community structure in the additive genetic (A), shared environmental (C), and dominant genetic (D) connectivity matrices, we applied a multi-resolution, multilayer community detection framework. For each variance component, a multiplex network was constructed with layers corresponding to different values of the resolution parameter ($\gamma$), enabling the assessment of modular organization across scales (see Section~\ref{sec:comm_det} and Figure~\ref{Flowchart_Hierarchy}). The resulting hierarchical community structures for the A, C, and D components are summarized in Figure~\ref{A_Community_render}.

Multislice community detection was then applied to each multiplex network to identify modules that are consistent within and across $\gamma$ values (Figure~\ref{comm_lines}).The top row of Figure~\ref{A_Community_render} displays the resulting dendrograms together with the corresponding reordered distance matrices for components A, C, and D. In these representations, colors in the dendrogram and blocks in the distance matrices denote groups of brain regions exhibiting similar community assignment patterns across resolutions. Darker blocks indicate sets of regions with highly stable co-assignment across scales, reflecting robust, component-specific modular organization.

Figure~\ref{A_Community_render} bottom row projects the final hierarchical modules for each component onto the cortical surface, revealing spatially distributed organizational patterns associated with additive genetics, shared environmental, and dominant genetics influences. Each component exhibits a distinct modular and hierarchical organization, indicating that genetic and environmental factors shape functional connectivity through partially overlapping yet component-specific organization into communities across multiple resolution scales.

\begin{figure}
	\centering
        \includegraphics[width=1\textwidth]{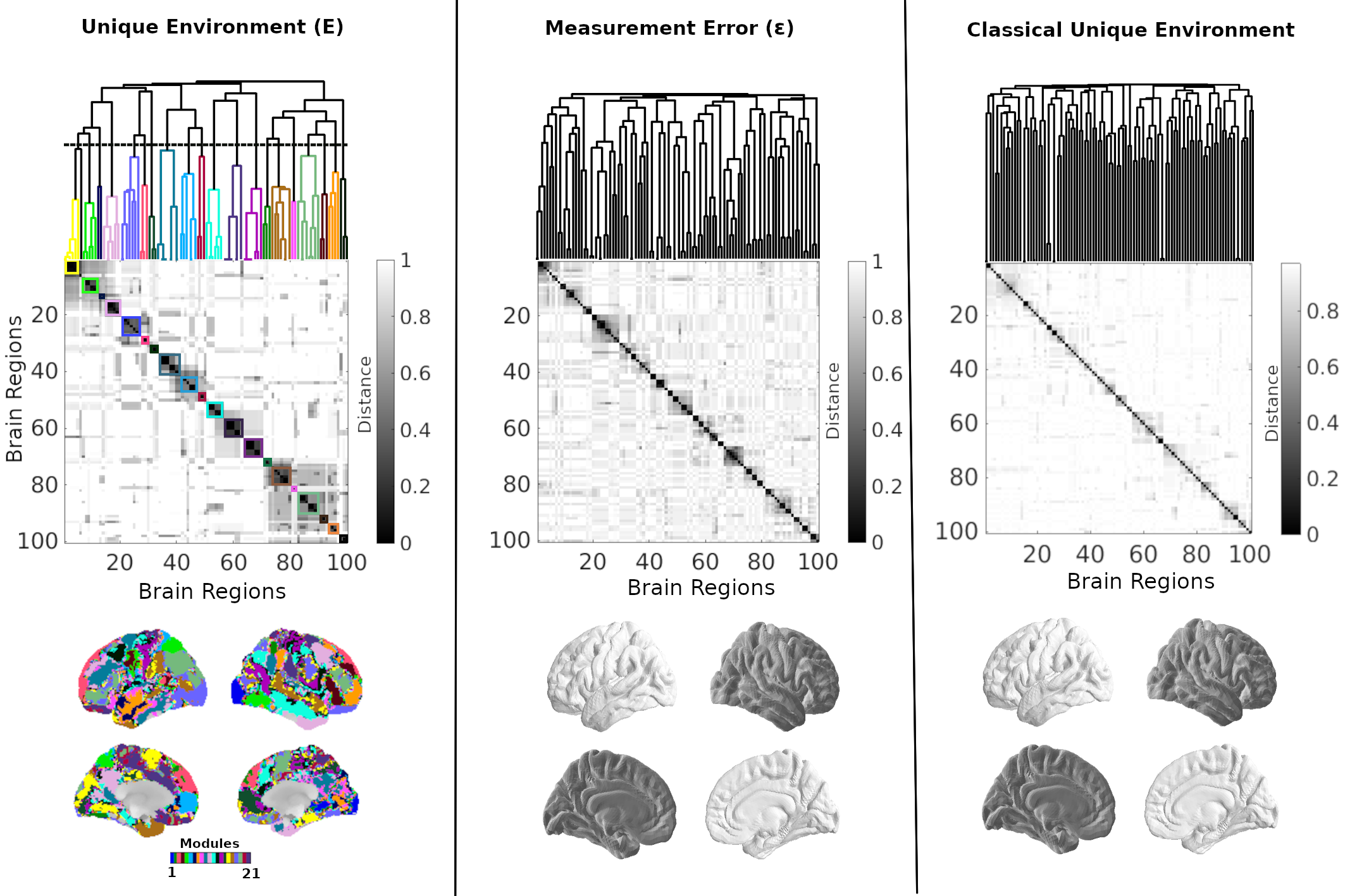}
	\caption{Hierarchical community structure of the refined unique environment (E), measurement error ($\mathcal{E}$), and classical unique-environment connectivity. For each component, the top panels display the dendrogram and corresponding distance matrix, illustrating the hierarchical clustering of brain regions based on the similarity of their modular organization across resolution scales. Notably, the classical unique-environment component exhibits patterns closely resembling those of measurement error, whereas the refined E component shows more coherent and structured modular organization. This contrast indicates that separating measurement error from the unique environment reveals meaningful environmental effects organized into communities that are otherwise obscured.
    }
	\label{E_Community_render}
\end{figure}

Figure~\ref{E_Community_render} illustrates the hierarchical community organization of the refined unique environment (E), measurement error ($\mathcal{E}$), and the classical unique-environment component. The refined E component exhibits a hierarchical structure that is clearly distinct from both measurement error and the classical formulation. In particular, the classical E term closely resembles $\mathcal{E}$, displaying fragmented, noise-like modular patterns, indicating that measurement error substantially dominates its structure. In contrast, the refined E component (left column) reveals coherent, spatially organized communities that are not apparent under the classical formulation. This separation indicates that decomposing the traditional unique-environment term into true environmental influences (E) and measurement error ($\mathcal{E}$) is essential, as otherwise the error component obscures meaningful environmental structure.

We next compared the partitions obtained by thresholding the hierarchical clustering (Figures~\ref{A_Community_render} and \ref{E_Community_render}) for each variance component. Results based on the Adjusted Rand Index (ARI; see Section~\autoref{sec:ARI}), summarized in Table~\ref{tabARI}, indicate that these partitions are distinct from one another and differ from canonical resting-state functional networks \citep{Yeo2011}. To further assess their relationship with canonical network organization, we quantified the extent to which each functional network is captured within single modules of each variance component (Table~\ref{tabFN_perc}). Overall, most functional networks are broadly distributed across multiple modules and components. Notable exceptions include the frontoparietal network, for which 46\% of regions are concentrated within a single module in the additive genetic (A) component, as well as the limbic (60\%) and visual (47\%) networks, which show greater concentration within single modules in the dominant genetic (D) component.

\begin{table}[]
\centering
\begin{tabular}{|p{2cm}|c|c|c|c|c|c|c}
\hline
Partitions & Yeo7 &  A & C & D & E \\
\hline
Yeo7 & 1 & 0.07 & 0.01 & 0.04 & 0.03\\
A & 0.07 & 1 & 0.01 & -0.02 & 0.009 \\
C & 0.01 & 0.01 & 1 & -0.016 & 0.006 \\
D & 0.04 & -0.02 & -0.016 & 1 & -0.005 \\
E & 0.03 & 0.009 & 0.006 & -0.005 & 1 \\

\hline
\end{tabular}
\caption{Pairwise similarity between component community partitions and functional network organization. Entries report the Adjusted Rand Index (ARI) between community labels derived from the additive genetic (A), shared environmental (C), dominant genetic (D), and unique environmental (E) components and the seven canonical functional networks. Diagonal entries equal unity by definition. Off-diagonal values quantify the degree of agreement between partitions, with values near zero indicating weak correspondence and negative values indicating agreement below chance.
}
\label{tabARI}
\end{table}

\begin{table}[]
\centering
\begin{tabular}{|p{4cm}|c|c|c|c|c|c|c}
\hline
FNs/Partitions & A(\%) & C(\%) & D(\%) & E(\%) & $\mathcal{E}$(\%) \\
\hline
Visual & 23.5 & 23.5 & 47.1 & 17.6 & 11.6 \\
Somatomotor & 42.8 & 35.7 & 21.4 & 35.7 & 21.4 \\
Dorsal Attention & 20 & 26.6 & 26.6 & 20 & 13.3  \\
Ventral Attention & 16.6 & 25 & 25 & 16.6 & 8.3  \\
Limbic & 20 & 40 & 60 & 20 & 20 \\
Fronto Parietal  & 46.1 & 30.7 & 23.1 & 15.3  & 15.3  \\
Default Mode Network  & 25 & 25 & 25 & 12.5  & 8.3  \\

\hline
\end{tabular}
\caption{For each functional network (rows), values indicate the maximum proportion of regions assigned to a single module within each connectivity component: additive genetics (A), shared environmental (C), dominant genetics (D), unique environmental (E), and measurement error ($\mathcal{E}$).
}
\label{tabFN_perc}
\end{table}

\section{Discussion}


In this study, we developed and applied an extended twin modeling framework to disentangle genetic, environmental, and measurement-related contributions to functional brain connectivity. By incorporating test--retest reliability into classical ACE and ADE models, we separated true environmental variance from measurement noise, yielding more interpretable and refined estimates of unshared environment in brain networks. Functional couplings were classified into four regimes—ACE$\mathcal{E}$, ADE$\mathcal{E}$, epistatic, and non-applicable—by retaining only those satisfying classical twin model assumptions. 

For each functional coupling in the connectome, we decomposed variance into additive genetics (A), shared environmental (C), dominant genetics (D), unique environmental (E), and measurement error ($\mathcal{E}$) components across resting-state and seven task fMRI conditions. To obtain reliable estimates for the majority of couplings, these component matrices were integrated using a minimum-error selection strategy, whereby, for each coupling, the estimate associated with the lowest $\mathcal{E}$ was retained. This procedure yielded unified A, C, D, and E matrices that capture the most reliable estimates across conditions. Finally, applying a multilayer, multi-resolution community detection framework revealed that genetic and environmental influences on functional connectivity are not randomly distributed, but instead organize into hierarchical, component-specific subnetworks across the cortex.

Table \ref{tabACDE} shows that demographic traits such as height and weight are predominantly influenced by additive genetic factors, whereas cognitive traits exhibit a more complex architecture involving additive, dominant, and epistatic effects. Notably, a substantial proportion of variance across cognitive traits is attributed to the non-shared environment ($E$), a component that also encompasses measurement error. These findings highlight that failing to disentangle measurement variability from true environmental effects can inflate estimates of $E$ and hinder the interpretation of environmental contributions. By incorporating test–retest reliability ($r_{TR}$), we were able to explicitly partition measurement noise ($\mathcal{E}$) from genuine non-shared environmental variance, thereby improving the interpretability and validity of both heritability and environmental estimates.

Recent work has emphasized the substantial impact of measurement error on heritability estimation, demonstrating that dissociating intra- and inter-subject variability can uncover genetic influences that are often obscured in conventional analyses. By explicitly modeling within-subject variability, these approaches show that a portion of the genetic signal may be masked by measurement noise, leading to underestimation of true heritability in complex phenotypes such as functional brain connectivity \citep{Ge2017, Chen2025}. Consistent with this perspective, Figure \ref{violin} illustrates the practical impact of extending the ACE/ADE framework to the ACE$\mathcal{E}$/ADE$\mathcal{E}$ models. Traditional formulations, in which $E = 1 - r_{MZ}$, systematically overestimate environmental contributions because this term also captures measurement error. By redefining $E = r_{TR} - r_{MZ}$ and explicitly modeling $\mathcal{E} = 1 - r_{TR}$, our framework separates true non-shared environmental variance from measurement error. This refinement yields more accurate estimates of environmental effects and indicates that a substantial fraction of what is typically attributed to “environment” in neuroimaging twin studies may instead reflect technical noise or instability in functional connectivity estimates.

Beyond improving variance decomposition, the measurement error matrix estimated for each fMRI condition constitutes a valuable result in its own right, as it enables both \textit{a priori} and \textit{a posteriori} assessment of the expected reproducibility of findings derived from functional connectivity data. Specifically, it provides a principled framework to quantify the reliability of associations or predictive models linking connectivity patterns to participant-level domains, such as phenotypic traits or cognitive measures. This has direct implications for widely used multivariate approaches, including connectome-based predictive modeling \citep{Shen2017}, partial least squares \citep{Guerrero2025, Mcintosh2004}, canonical correlation analysis, and tensor-based methods such as CP decomposition \citep{Carvalho2025}. From an \textit{a priori} perspective, the measurement error matrix can be used to restrict analyses to functional couplings with lower estimated error, thereby improving the stability of downstream inferences. Complementarily, from an \textit{a posteriori} perspective, the uncertainty associated with model outputs—such as component loadings or connectivity weights—can be quantified using the same framework, enabling a more rigorous assessment of the reliability of derived solutions.

Figure \ref{flowchart} summarizes the classification of functional couplings under the extended twin modeling framework. By applying conditional thresholds on monozygotic ($r_{MZ}$) and dizygotic ($r_{DZ}$) twin correlations, each coupling was assigned to the ACE$\mathcal{E}$ or ADE$\mathcal{E}$ formulations, or excluded due to violation of model assumptions or evidence of epistatic effects. Our framework extends the classical twin model by explicitly incorporating test–retest reliability ($r_{TR}$), enabling the separation of true environmental variance from measurement noise ($\mathcal{E}$). The resulting formulations provide corrected estimates of additive (A), dominant (D), shared (C), and unique environmental (E) contributions, subject to the updated constraint $A + C + E + \mathcal{E} = 1$ or $A + D + E + \mathcal{E} = 1$

This extension is essential for improving interpretability in analyses of functional connectomes. By restricting inference to couplings that satisfy classical twin model assumptions and explicitly accounting for measurement error, the ACE$\mathcal{E}$/ADE$\mathcal{E}$ framework enables a more accurate quantification of the relative contributions of genetic and environmental factors to brain connectivity. Moreover, the epistasis regime ($r_{MZ} > 4r_{DZ}$) captures couplings consistent with non-additive gene–gene interactions, underscoring the complexity of genetic influences beyond additive and dominant effects. Together, this structured classification provides the foundation for subsequent analyses, including the spatial mapping and hierarchical organization of heritable connectivity patterns.



Figure \ref{variance_components} shows that the majority of functional couplings follow the ACE$\mathcal{E}$ model, indicating that additive genetic and shared environmental effects account for a substantial portion of functional connectivity variability across both resting-state and task conditions. A smaller but consistent fraction of couplings is better explained by the ADE$\mathcal{E}$ model, reflecting contributions from dominant genetic interactions. The presence of epistatic and non-applicable couplings further underscores the complexity and heterogeneity of heritability across the connectome. Importantly, the high consistency between test and retest sessions supports the robustness of these patterns, and averaging across sessions increases the fraction of couplings that satisfy model assumptions, thereby improving the reliability of downstream analyses.

These results also highlight the critical importance of appropriately determining when ACE, ADE, or neither model should be applied. Systematically imposing the ACE model, without verifying the underlying correlation structure, can lead to biased variance decomposition. In particular, when the ACE assumptions are violated, the shared environment term (C) can become negative, and because the total variance is constrained to sum to one, this artifact artificially inflates estimates of additive genetic effects (A). Such mis-specification can lead to overestimation of heritability and misinterpretation of the relative contributions of genetic and environmental factors \citep{Calhoun2024, Colclough2017}. By explicitly enforcing model selection criteria based on the relationships between $r_{MZ}$ and $r_{DZ}$, it is possible to avoid these biases and ensures that variance components are interpreted within their appropriate regimes.


The observed variation in the fraction of non-applicable couplings across resting-state and task fMRI conditions reflects contributions from both scan length and fMRI condition. Consistent with prior work demonstrating the impact of acquisition duration on the reliability of functional connectivity estimates \citep{Birn2013, Ooi2025, Jarrahi2021, Wahab2022}, shorter scans were associated with a substantial reduction in the proportion of analyzable couplings. However, this effect cannot be attributed solely to scan length. As shown in Figure~\ref{na_scan}, the fraction of non-applicable couplings decreased with increasing scan length but also differed systematically between resting-state and task conditions. Multiple linear regression analysis further confirmed that both scan length and condition are significant predictors of this variability. Across most tasks, the proportion of non-applicable couplings was lower than in resting state at comparable scan durations, suggesting that task engagement may enhance the stability of connectivity estimates. The motor task constituted a notable exception, exhibiting a higher fraction of non-applicable couplings than resting state at similar scan lengths, potentially reflecting its comparatively lower cognitive demands. Together, these findings indicate that acquisition duration and brain state jointly shape the reliability of inter-individual connectivity estimates.

To address this limitation, we developed a minimum-error integration strategy that combines information across all fMRI conditions to maximize reliability. For each functional coupling, the estimate corresponding to the condition with the lowest measurement error ($\mathcal{E}$) is retained, thereby accounting for scan-specific variability and enhancing the stability of variance component estimates. This approach enables the aggregation of data across conditions with heterogeneous reliability, yielding a more robust and interpretable representation of the underlying genetic architecture of functional connectivity.

Applying a multilayer, multi-resolution community detection framework to the integrated A/C/D/E matrices revealed a hierarchical modular organization of genetic and environmental influences on functional connectivity. For each resolution parameter ($\gamma$), a consensus matrix was computed and treated as a layer within a multiplex network, enabling the identification of modules that are consistent across scales. The resulting distance matrices and dendrograms summarize inter-regional similarity in community assignment patterns, revealing a hierarchical clustering that is subsequently projected onto the cortical surface. Together, these findings demonstrate that heritable and environmental effects on the functional connectome are organized into spatially distributed, component-specific, and hierarchically structured communities. Cortical projections further indicate that these modules occupy distinct yet partially overlapping brain regions. The observed hierarchical organization suggests that genetic and environment influences operate across multiple spatial scales, spanning both localized subnetworks and distributed large-scale systems.

The present work has several limitations that warrant consideration. First, the sample is restricted to a young adult cohort; extending this framework to older populations may reveal age-related shifts in the balance between genetic and environmental influences \citep{Briley2013, Haworth2010} on functional connectivity. Second, although measurement noise was explicitly modeled using test–retest reliability, residual sources of variability—such as physiological artifacts and preprocessing-related differences—may still contribute to the estimated error term ($\mathcal{E}$). Third, the relatively modest sample size limits statistical power, particularly for detecting shared environmental and epistatic effects. Future studies with larger cohorts and longer scan durations may yield more precise variance component estimates and increase the detectable contribution of true environmental influences.

At present, the Human Connectome Project (HCP) provides an openly available collection of resting-state fMRI data from monozygotic and dizygotic twins with well-characterized family structure. As larger and more demographically diverse twin datasets become available, future studies will be able to validate and extend the present findings. In addition, diffusion MRI offers a complementary avenue to characterize heritable aspects of white-matter architecture and, when integrated with functional connectivity, may provide a more comprehensive view of the structural substrates underlying genetic and environmental influences. Incorporating detailed demographic, behavioral, and cognitive measures will further enable the investigation of associations between brain connectivity and phenotypic traits, including the assessment of shared genetic contributions.

The proposed framework is also readily extensible to multimodal data integration. For example, combining structural and functional MRI within the same modeling framework would enable the characterization of heritability across both brain architecture and dynamics. In this context, the multilayer community detection approach could be extended to identify shared genetic patterns spanning structural and functional domains, providing a unified view of how genetic influences shape both brain organization and activity.


In summary, this work introduces an extended twin modeling framework that explicitly separates measurement error from environmental variance, enabling more accurate and interpretable estimates of heritability and environment in functional brain connectivity. By integrating data across multiple fMRI conditions and characterizing genetic and environmental organization using multilayer community detection, we show that these influences are spatially distributed and hierarchically structured across the cortex. These findings highlight the complex genetic and environmental architecture underlying brain network organization and provide a principled foundation for future studies linking specific brain connectivity patterns, cognition, and behavior to genetic and environmental factors.

\newpage
\bibliographystyle{apacite}
\bibliography{NETNbibsamp}

\acknowledgments
This work was supported by NIH CTSI CTR EPAR2169, NIH R21 AA029614, NIH R01 AA029607, Indiana Alcohol Research Center P60AA07611, NINDS R01NS126449 and NINDS R01NS112303

\section{Supporting Information}
\setcounter{figure}{0}
\renewcommand{\thefigure}{S\arabic{figure}}

\begin{figure}[H]
	\centering
        \includegraphics[width=1\textwidth]{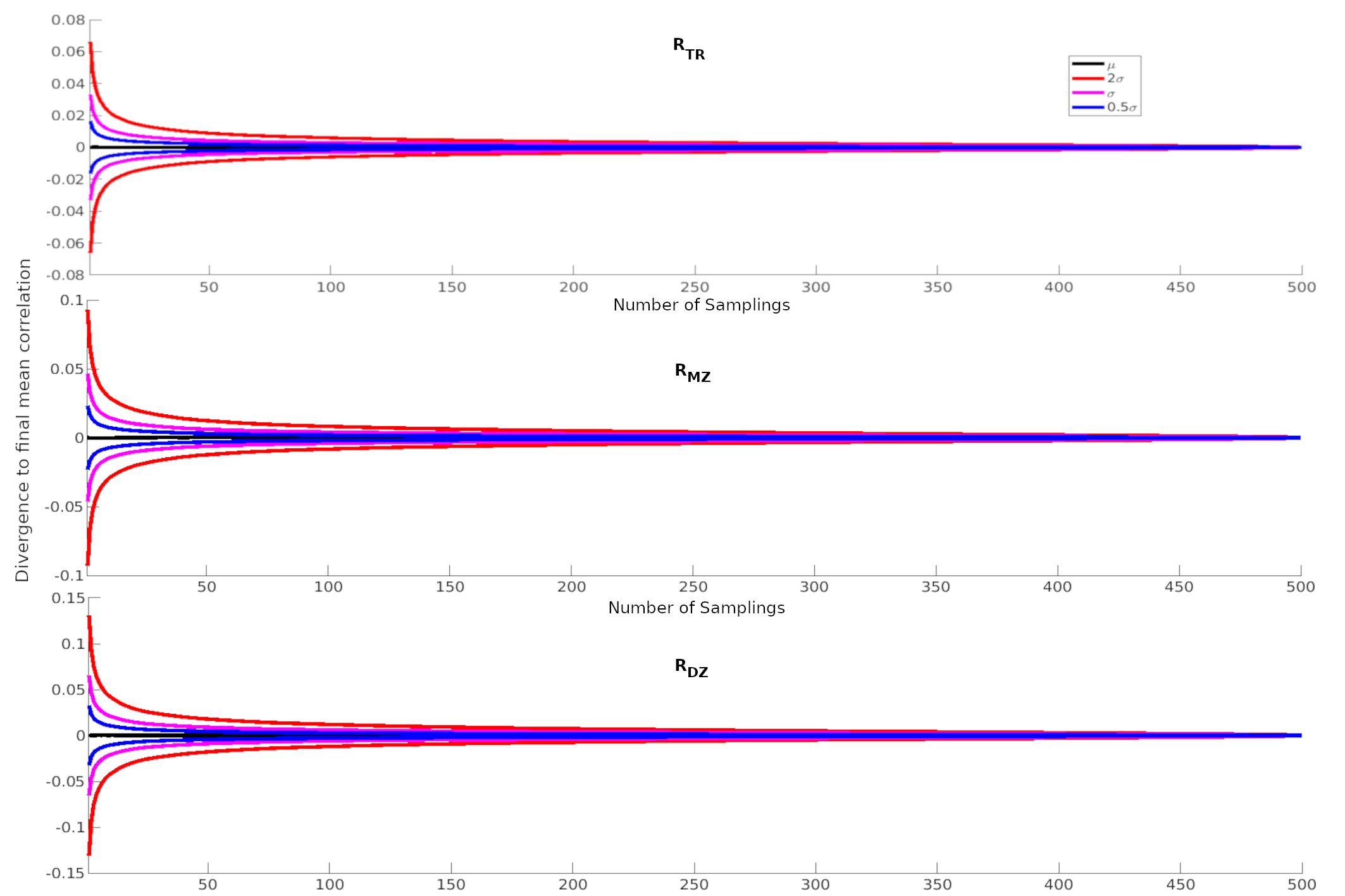}
	\caption{Convergence of correlation estimates across repeated random samplings for test–retest, monozygotic (MZ), and dizygotic (DZ) twin pairs. Each panel shows the divergence of sampled correlation coefficients from the mean correlation as the number of samples increases. The black line represents the mean ($\mu$) across all iterations, while colored lines indicate $\pm 0.5\sigma$ (blue),  $\pm \sigma$ (magenta), and $\pm 2\sigma$ (red) bounds. As the number of samplings increases, the variability of the estimates diminishes and the curves converge toward stable mean correlations, demonstrating reliability of the sampling-based correlation estimation procedure. }
	\label{convergence_corr}
\end{figure}

\begin{figure}[H]
	\centering
        \includegraphics[width=1\textwidth]{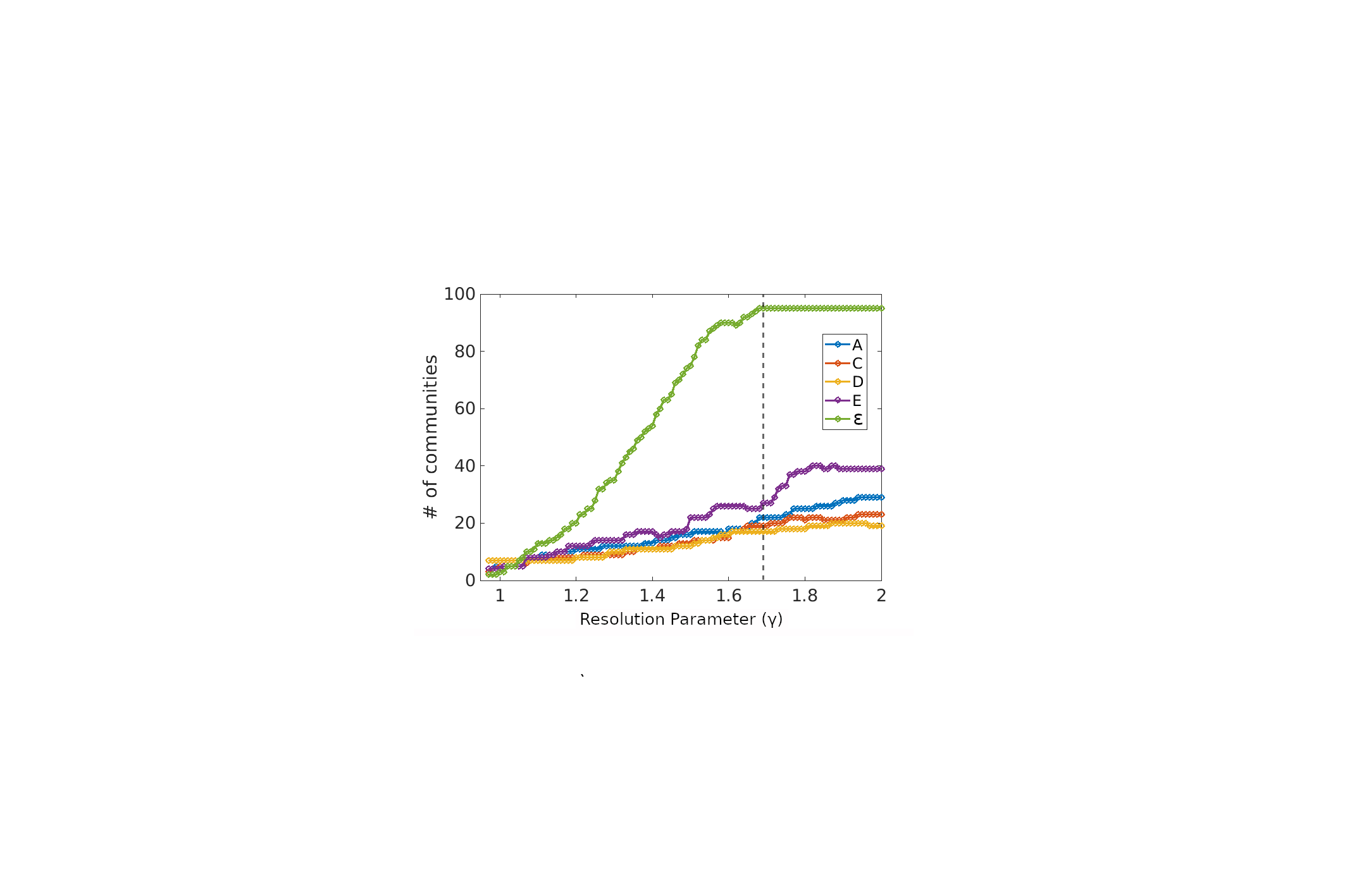}
        \vspace{-4cm}
	\caption{Number of detected communities for each component (A, C, D, E, and $\mathcal{E}$) across the resolution parameter ($\gamma$). For the measurement error ($\mathcal{E}$) matrix, the number of communities stabilizes at $\gamma=1.69$ (vertical dashed line), which we use as the maximum $\gamma$ value when building the multilayer network. The minimum $\gamma$ value in the range is chosen as the point where all components first show at least two communities.}
	\label{num_comm}
\end{figure}

\begin{figure}[htbp]
	\centering
        \includegraphics[height=0.7\textheight, keepaspectratio]{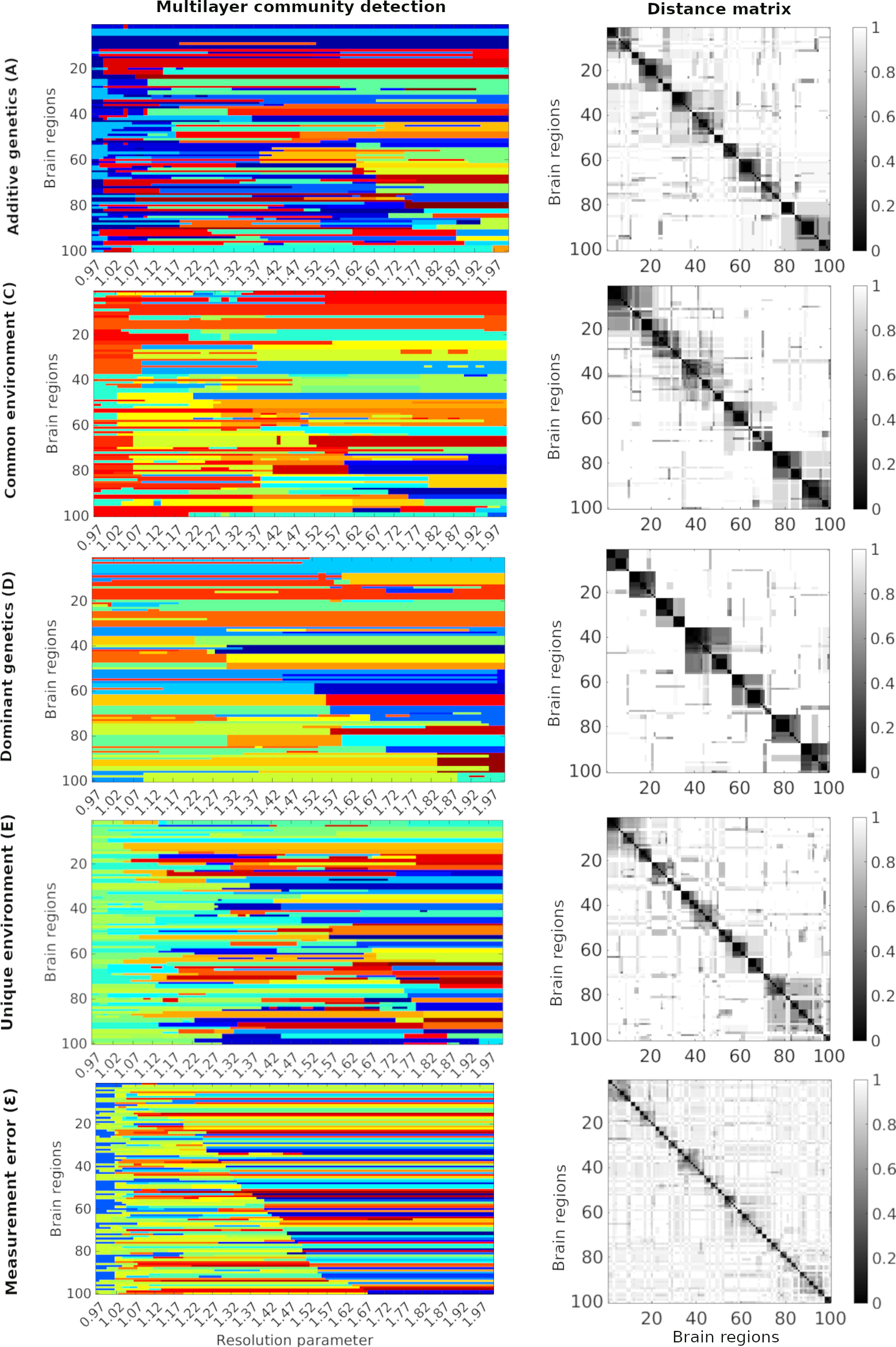}
	\caption{Multilayer community detection for the additive genetics (A), common environment (C), dominant genetics (D), unique environment (E), and measurement error ($\gamma$) multiplex networks (left). For each component, brain region community (rows) are shown across the entire $\gamma$ range, illustrating how modular structure evolves with increasing resolution (columns). The panels on the right display the associated distance matrices, reordered according to hierarchical clustering, which identify sets of regions that exhibit similar modular trajectories across the $\gamma$ range, revealing the hierarchical organization underlying each component.}
	\label{comm_lines}
\end{figure}

\begin{figure}
	\centering
        \includegraphics[width=1\textwidth]{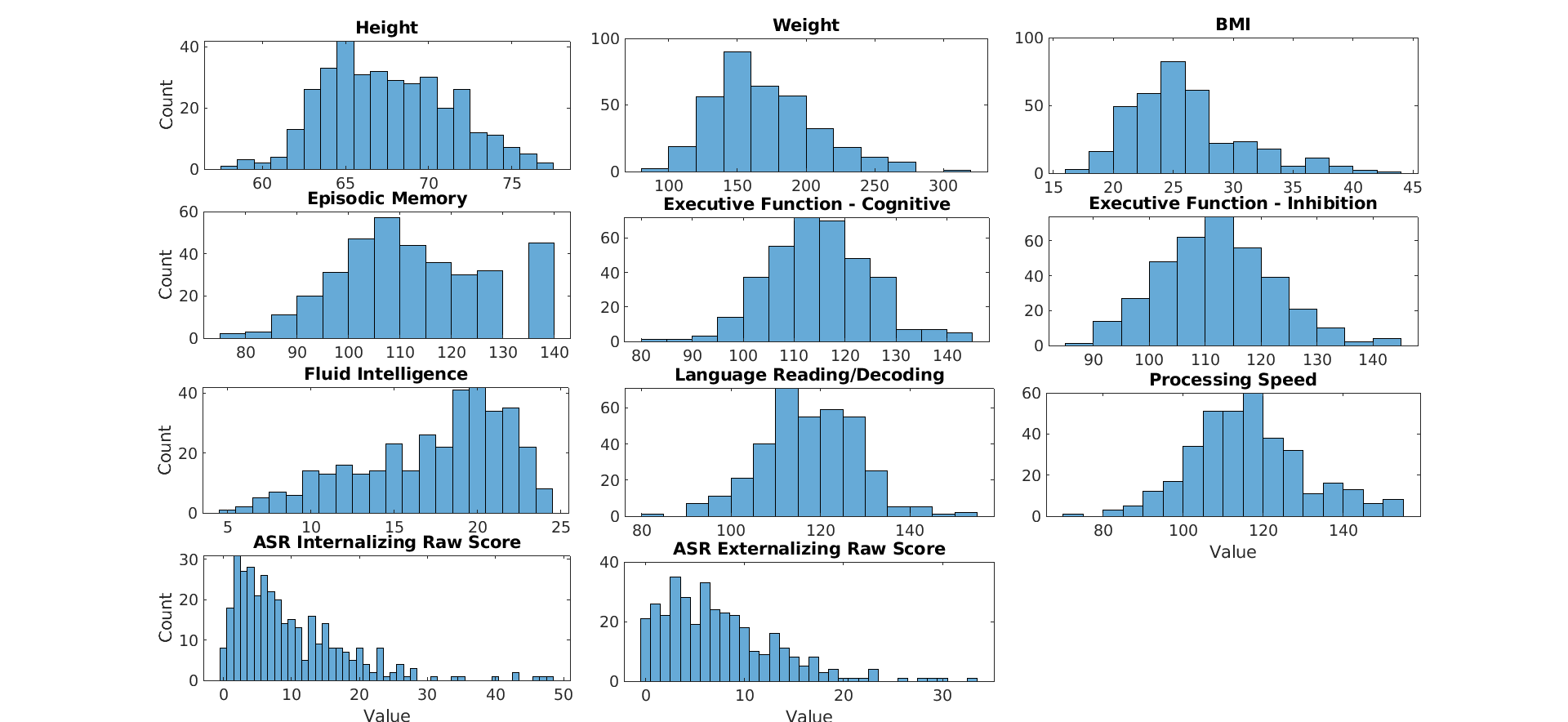}

	\caption{Distribution of demographic, cognitive, and behavioral traits across the full sample.
Histograms illustrate the distribution of each trait, including height, weight, BMI, cognitive performance (episodic memory, executive function—cognitive flexibility and inhibition, fluid intelligence, language reading/decoding, and processing speed), and behavioral measures (ASR internalizing and externalizing raw scores).}
	\label{trait_histogram}
\end{figure}

\end{document}